\documentclass{article}

\usepackage{arxiv}

\usepackage[utf8x]{inputenc} 
\usepackage[T1]{fontenc}    
\usepackage{hyperref}       
\usepackage{url}            
\usepackage{booktabs}       
\usepackage{amsfonts}       
\usepackage{nicefrac}       
\usepackage{microtype}      
\usepackage{graphicx}
\usepackage[numbers]{natbib}
\usepackage{doi}
\usepackage{balance} 
\usepackage{multirow}
\usepackage{amsthm}
\usepackage{amsmath}
\usepackage{enumitem}
\usepackage{xcolor}
\usepackage[english]{babel}

\usepackage{subfiles}
\usepackage[resetlabels,labeled]{multibib}
\usepackage{soul}
\usepackage{float}

\newcommand{\mev}{\texttt{MEV}}
\newcommand{\mevg}{\texttt{MEV-Geth}}
\newcommand{\mevb}{\texttt{MEV-Boost}}
\newcommand{\pbs}{\texttt{PBS}}
\newcommand{\mevl}{\texttt{MEV-Blocker}}
\newcommand{\meve}{\texttt{MEV-Explore}}
\newcommand{\suave}{\texttt{SUAVE}}
\newcommand{\mevm}{\texttt{MEVM}}
\newcommand{\evm}{\texttt{EVM}}
\title{MEV Ecosystem Evolution From Ethereum 1.0}

\author{
    {Rasheed}\\Machine Learning Lab \\ International Institute of Information Technology,\\ Hyderabad, India\\\texttt{\href{mailto:}{}mohammad.ahmed@research.iiit.ac.in}
\And 
    {Yash Chaurasia}\\Machine Learning Lab \\ International Institute of Information Technology, \\Hyderabad, India\\\texttt{yash.chaurasia@research.iiit.ac.in}
\And 
    {Parth Desai}\\Machine Learning Lab \\ International Institute of Information Technology, \\Hyderabad, India\\\texttt{parth.desai@research.iiit.ac.in}
\And 
    {Sujit Gujar}\\Machine Learning Lab \\ International Institute of Information Technology, \\Hyderabad, India\\\texttt{sujit.gujar@iiit.ac.in}
}

\date{}

\newcommand{\subsubsubsection}[1]{\paragraph{#1}\mbox{}\\}
\begin{document}
\maketitle
\begin{abstract}
    Smart contracts led to the emergence of the decentralized finance (DeFi) marketplace within blockchain ecosystems, where diverse participants engage in financial activities.
In traditional finance, there are possibilities to create values, e.g., arbitrage offers to create value from market inefficiencies or front-running offers to extract value for the participants having privileged roles. Such opportunities are readily available -- searching programmatically in DeFi. It is commonly known as Maximal Extractable Value (\mev) in the literature. In this survey, first, we show how lucrative such opportunities can be. Next, we discuss how protocol-following participants trying to capture such opportunities threaten to sabotage blockchain's performance and the core tenets of decentralization, transparency, and trustlessness that blockchains are based on. Then, we explain different attempts by the community in the past to address these issues and the problems introduced by these solutions. Finally, we review the current state of research trying to restore trustlessness and decentralization to provide all DeFi participants with a fair marketplace.

The study is structured as follows.

\end{abstract}
\tableofcontents
\newpage
\section{Introduction}
\label{sec:introduction}

\subsection{Blockchain}
\label{ssec:blockchain:introduction}
Bitcoin~\cite{nakamoto} introduced us to \emph{blockchain}, a distributed ledger technology (DLT) that allows participants to commit a block of transactions as per a consensus mechanism. Each new block contains the previous block's hash, indicating the blockchain's previous state.

Ethereum introduced the \emph{Ethereum Virtual Machine} (\evm), allowing a Turing complete set of instructions to be written and interpreted on the blockchain. This allows users to write \emph{smart contracts}, code that lives and evolves on the blockchain. Smart contracts gave rise to various \emph{Decentralized Applications} (dApps). This report focuses on the problems propped up by the rise of \emph{Decentralized Finance} (DeFi) dApps that deal with cryptocurrencies.

\subsection{Background}
\label{ssec:background:introduction} 

\subsubsection{Centralized Exchange}
\emph{Centralized Exchanges} (CEX) are firms that facilitate and coordinate trading at a large scale. They operate as intermediaries between buyers and sellers, acting similarly to banks where users trust the exchange to handle their assets securely. While stock and metal exchanges were significant entities in the traditional economy, Coinbase, Binance, etc., are some of the prominent entities in the crypto economy.

CEXs process transactions more efficiently during high loads, offering fast execution of orders. But CEXs are managed by centralized parties and are often prone to a single failure, resulting in huge losses \cite{MtGox}. Centrally-operated cryptocurrency exchanges (CEXs) contradict the decentralized principles inherent in blockchain technology. Further, there is a lack of transparency and censorship resistance as CEXs can misuse or mismanage user funds or engage in practices like wash trading. Thus, it has led to the development of decentralized exchanges.

\subsection{Decentralized Finance}
\label{ssec:defi:introduction}
\emph{Decentralized Finance} (DeFi) is an alternative autonomous financial system that functions transparently and permissionless. DeFi cuts out the middleman in finance. DeFi uses smart contracts and self-executing blockchain agreements to facilitating lending, borrowing, trading, and other financial activities without relying on banks or traditional institutions.

\subsubsection{Decentralized Exchange}
\emph{Decentralized Exchanges} (DEX) are a key part of DeFi. These are marketplaces built upon blockchain systems to provide decentralized trading. Due to decentralized settlements, DEXs are transparent and offer greater security. Users can control their funds and withdraw safely even when DEXs stop working. They offer censorship resistance. 

DEX runs automated algorithms instead of the conventional approach of acting as a financial intermediary between buyers and sellers. These algorithms are run in the form of smart contracts. These contracts can be executed automatically when certain conditions are met, allowing users to interact directly and enabling peer-to-peer lending. However, DEXs are not fast and scalable compared to CEXs.

\subsubsection{Profitable Opportunities on Blockchain}
\emph{Decentralized Finance} (DeFi) creates many pure, profitable trading opportunities, such as arbitrage and liquidations on automated markets run by smart contracts. Further, the ordering of transactions is controlled by miners or validators, which can change for each block and the transaction order. The validators can choose to optimize the ordering of transactions to extract value and profit for themselves. 

Consider an arbitrage opportunity where Token $X$ is listed on $DEX_1$ at a price $A$ and $DEX_2$ at $B$. Let $A < B$. A bot that is constantly monitoring the mempool identifies this price discrepancy. It quickly submits a transaction to the mempool to buy Token $X$ from $DEX_1$ at a price $A$ and sell it at $DEX_2$ at $B$, earning $B - A$ per token. The bot can continue until the updated prices on DEXs adjust such that $A' = B'$. To incentivize miners to include this transaction in the next block, the bot offers a higher gas price than other pending transactions.

Such opportunities are not limited to DeFi but also can be found in other applications such as \emph{non-fungible tokens} (NFTs). During NFT drops, when many buyers compete to purchase NFTs, miners can reorder transactions and profit themselves. Informally, the maximum profit from such opportunities is called the \emph{maximal extractable value} (\mev). 
\subsection{What is \mev?}
\label{ssec:what_is_mev:introduction}
As defined by Ethereum \cite{ethereummev}, \emph{Maximal Extractable Value (\mev) refers to the maximum value that can be extracted from block production in addition to the standard block reward and gas fees by including, excluding, and changing the order of transactions in a block.}

Before \emph{The Merge} (the event of Ethereum's shifting its underlying consensus mechanism to a more energy-efficient one, \mev\ referred to \emph{Miner Extractable Value}. Miners were responsible for block building via mining and block production. Miners could order transactions and, thus, leverage their powers to gain more revenue. After The Merge, Ethereum introduced new roles for block building, and miners were referred to as validators/proposers. We discuss this in detail in Section \ref{sec:preliminaries}.

The theoretical maximum value that can be extracted exceeds the actual value extracted. Flashbots call the detected extracted value as \emph{Realised Extractable Value} (REV) \cite{quantifyingrev}. However, we use the more commonly used term \mev\ in this paper.

\subsection{Why is \mev\ Significant?}
\label{ssec:mev_significance:introduction}
\begin{figure}
\includegraphics[width=\textwidth]{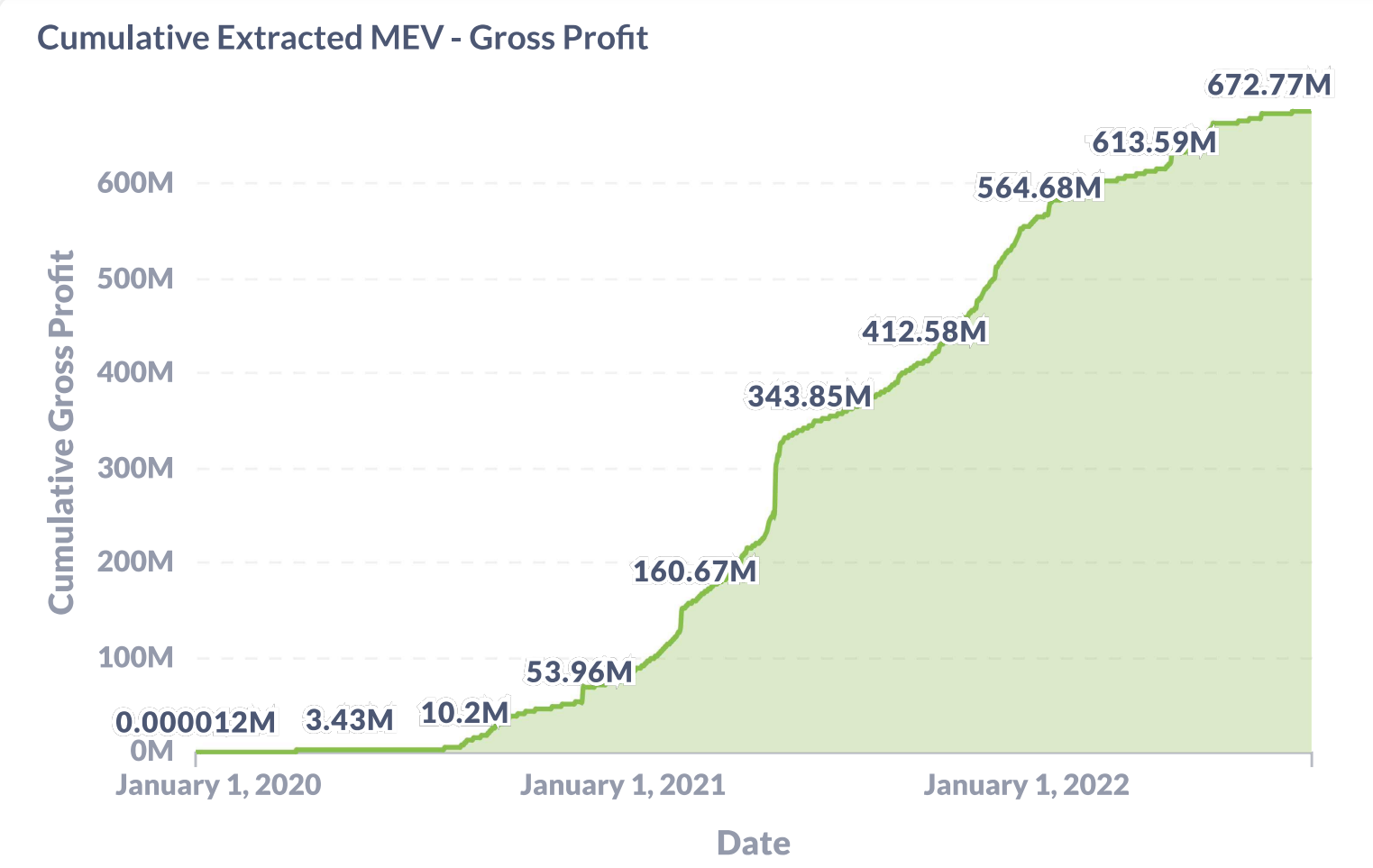}
\caption{Cummulative gross profit  in dollars extracted pre-merge\cite{DBLP:journals/corr/abs-1904-05234}}
\label{fig:CGP}
\vspace{-4mm}
\end{figure}

\begin{figure*}
\includegraphics[width=\linewidth]{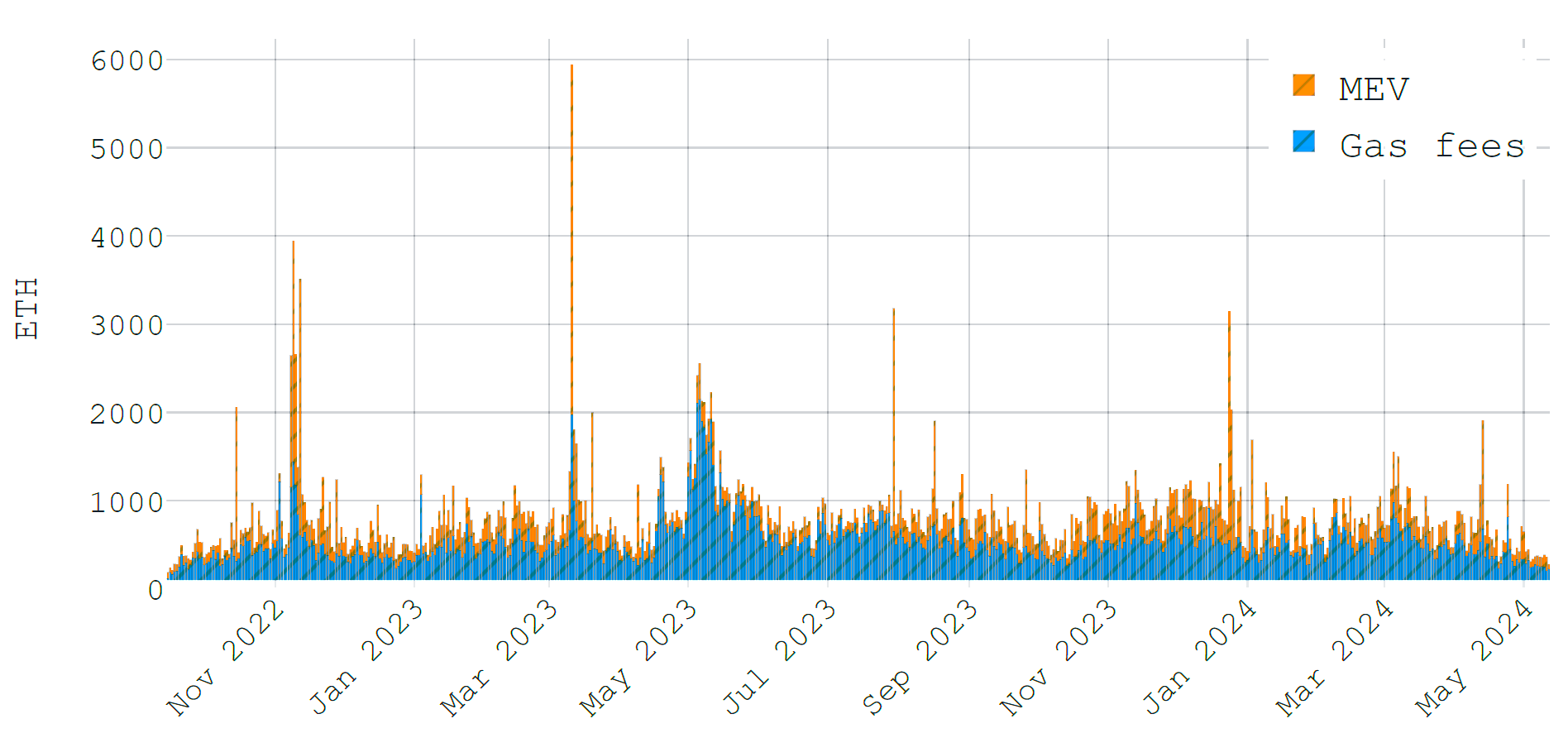}
\caption{\mev\ revenue and gas fee post-merge \cite{mevBoostDash}}
\label{fig:MEVPOST}
\vspace{-4mm}
\end{figure*}

\begin{figure}
\centering
\includegraphics{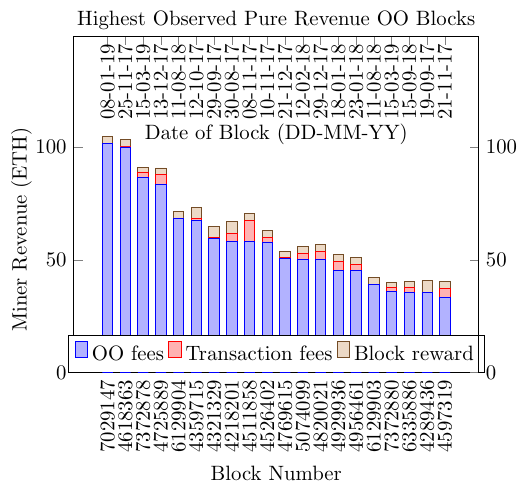}
\caption{Ethereum blocks in which \mev\ value dominates both block rewards and transaction fees as mentioned by FlashBoys Team ~\cite{DBLP:journals/corr/abs-1904-05234}}\label{fig:hotoo}
\vspace{-4mm}
\end{figure}

The cumulative gross profit of the \mev\ earned in dollars as of Jan 2022, observed by Flashbot \meve v1~\cite{mevcgp}, is shown in Figure~\ref{fig:CGP}. Figure \ref{fig:MEVPOST} shows the \mev\ Revenue in ETH and the gas fee post-merge. Both graphs affirm that \mev\ opportunities were consistently available in the Ethereum network. While extracted, \mev\ might not always be high; there have been instances of \mev\ revenue surpassing the block rewards and transaction fees. Consider Figure~\ref{fig:hotoo}, which shows Ethereum blocks where \mev\ extracted by ordering transactions is higher than block rewards and transaction fees. Here, Ordering Optimization (OO) fees reflect implicit fees a miner can reap by leveraging their control of a consensus epoch.  \mev\ represents a significant economic incentive for miners or validators in blockchain networks. They can optimize revenue by strategically ordering transactions and extracting value from trades, liquidations, or other DeFi activities.

\mev\ directly impacts economic incentives, market dynamics, security, protocol design, and governance, shaping the evolution and sustainability of decentralized systems. As rational players, participants involved directly or indirectly are incentivized to act strategically, which might hamper the security and consensus stability of the network. 

\subsection{Issues with \mev}
\label{ssec:issues_with_mev:introduction}
While \mev\ extraction balances the market inefficiencies and brings the right value to assets \cite{ethereummev}, \mev\ has negative externalities. Automated trading bots quickly capture these limited and ephemeral \mev\ opportunities. In the race to capture the \mev, nodes compete with each other by bidding higher transaction fees, resulting in a repeated game among network participants, who compete to claim the opportunity. This leads to network congestion and bloating of transaction mempool.

During \mev\ extraction, the transaction fees paid to the miner increase drastically. Incentivized by the transaction fee, miners tend to pick up only high-valued transactions, causing stagnation in low-valued transactions. Further, though blockchains aim at decentralization, \mev\ extraction becomes centralized as miners fully control transaction ordering. Censorship in \mev\ extraction becomes relatively more significant since the miner can exploit users' transactions by replacing them with their transactions, resulting in a loss for the users. In such cases, users lose \mev\ value and pay the transaction fee when it gets included later (when the \mev\ opportunity isn't available anymore).\\

Lastly, entities may wield disproportionate influence over others' decision-making processes, such as the role of top builders in the current scenario (more about this in section \ref{ssec:block_auctions}. They can act strategically without any protocol violation and incentivize participants trying to capture \mev\ to favor their interests, potentially undermining the decentralized nature of blockchain networks and resulting in the centralization of power.

\subsection{Objective of the Survey}
\label{ssec:objectives:introduction}
This survey seeks to delve into the evolution of \mev. In particular, this paper delves into the nuanced shifts in transaction flow and incentive structures within the \mev\ ecosystem, shedding light on their implications for the Ethereum network. Moreover, it delves into ongoing research endeavors in this domain and provides critical insights into the current proposals under scrutiny within the DeFi community.

\subsection{Our Approach to Unraveling the \mev\ landscape}
\label{ssec:approach:introduction}
We discuss various stages of MEV evolution with a brief introduction covering the motivation and various parties/components involved. Further, we scrutinize the transaction flow within the system, identifying and dissecting prevalent issues therein. 

\begin{figure}[H]
    \centering
    \includegraphics[scale=0.8]{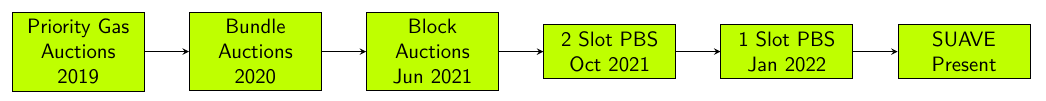}
    \caption{Timeline of \mev}
    \label{fig:Timeline MEV}
\end{figure}

At each stage of \mev\ evolution, we endeavor to furnish publicly available observations reported by various organizations. Through this approach, we aspire to unveil the ramifications of \mev\, offering valuable insights into its true impact. Figure \ref{fig:Timeline MEV} shows the evolution of \mev\ with time.

\subsection{Organization of Paper}
\label{ssec:paper_organization:introduction}
The rest of the paper is organized as follows: Section \ref{sec:preliminaries} discusses preliminaries of blockchain and DeFi. Section \ref{sec:mev_landscape_1} discusses \mev\ landscape in Ethereum 1.0, and Section \ref{sec:mev_landscape_2} discusses \mev\ landscape in Ethereum 2.0. Further Section \ref{sec:private_order_flow} talks about private order flow and its impact on the \mev\ extraction. Finally, Section \ref{sec:current_development} sheds light on the \mev\ landscape developments and explores the impact of the latest proposals.

\section{Preliminaries}
\label{sec:preliminaries}
This section serves as an introduction to some essential concepts. These concepts lay the foundation for understanding the backdrop of this survey report.
\subsection{Blockchain}
\label{sec:blockchain}
 \emph{A blockchain is a distributed ledger with growing lists of records (blocks) that are securely linked together via cryptographic hashes.} \cite{enwiki:1220180266:blockchain}. This information is stored in blocks, where each block’s information is verifiable by every participating computer. It’s designed to have decentralized management instead of the traditional hierarchical systems we’re familiar with. A dispersed structure like the blockchain helps to ensure trust, validity, and usability.

 A block in a blockchain primarily consists of:
\begin{itemize}
    \item List of transactions - Valid transactions signed by the sender are accepted once they become a part of the accepted block.
    \item Hash of the previous block - allows the block producer to commit to the blockchain history it considers valid inductively. This renders blockchains immutable.
\end{itemize}

A consensus mechanism dictates which proposed blocks are considered valid.
Blockchain participants broadcast valid blocks and proposed transactions they hear to other connected network participants. We discuss these in detail in the following sections.


\subsubsection{Transaction}
A ledger is a collection of transactions from one account to another. Similarly, a blockchain is a collection of publicly accepted transactions. A transaction can be a token ownership transfer, as in Bitcoin~\footnote{Apart from transferring an owner's UTXOs to a public key, other types of transactions are also possible on Bitcoin using Script. In the case of a smart contract transaction on Ethereum, SegWit, and Taproot updates further offer a wider variety of transactions.}, or even a valid \evm\ state change.

In most blockchains, transaction creators, aka users, whom we refer to as \emph{users}, can pay transaction fees to speed up their transaction's inclusion.

\subsubsection{Transaction Mempool}
A transaction signed by a user is broadcasted to the network for inclusion in a future block. Every miner/block producer validates the transactions it receives and, if valid, adds them to a list of transactions to be added to the blocks. This list of transactions waiting to be added is referred to as \emph{transaction mempool}. It is a staging area for transactions, and some blockchains allow transactions to be canceled or replaced while in the mempool.

\subsubsection{Publishing Blocks}
The success of blockchain relies on the number of participants volunteering to pick valid transactions up from the mempool and produce a valid block. Most blockchains offer a pre-defined reward to block publishers for their efforts. Each blockchain specifies a mechanism to decide a participant's eligibility to publish a block. Such participants are called \emph{miners} in PoW-like consensus mechanisms and \emph{validators} in PoS-like consensus mechanisms.

Blockchains also specify the desired block size and block publishing rate of the network. In Bitcoin, one block is published in around $10$ minutes. In Ethereum, one block is published every $12$ seconds. There is a trade-off between block size, block publishing rate, and the network's security, limiting the throughput blockchains can achieve.

\subsubsection{Consensus}
Participants of a distributed system need a mechanism to achieve consensus on updates proposed in the network. Bitcoin~\cite{nakamoto} uses \emph{proof-of-work} (PoW) (also known as Nakamoto consensus) as its consensus mechanism. Ethereum recently shifted from proof-of-work to \emph{proof-of-stake} (PoS). We discuss both PoW and PoS in the following sections.

\subsubsubsection{Proof-of-Work}
One must solve a puzzle to publish a block in PoW, proving they have done computationally intensive work. This renders Sybil attacks useless. While any participant can spin up new identities, they are limited in their total computational power. As discussed in~\cite{bitcoin-book}, the following properties are essential for a puzzle to be used to achieve consensus:
\begin{itemize}
    \item Easy-verifiability - The solution to the puzzle must be easily verifiable so that everyone can verify it.
    \item Proportional advantage - The probability of mining a block should be proportional to the fraction of the participant's computational power. Thus, unlike traditional voting-like mechanisms, there's no advantage in generating more identities.
    \item Adjustable difficulty - The difficulty of finding a solution should be adjustable to maintain a pre-defined block publishing rate.
    \item Non-reusable - A solution for a block shouldn't be usable for any other block.
\end{itemize}

\subsubsubsection{Miner}

There are three types of participants in the blockchain network:
\begin{itemize}
    \item \emph{Light nodes}: Participants who do not invest a lot of computational power and only store and retrieve (from full nodes) the information they need to verify their own transaction status.
    \item \emph{Full nodes}: Participants store the entire blockchain and verify the validity of all transactions and blockchain state changes.
    \item \emph{Miner}: Full nodes that deploy their computational resources to mine new blocks according to the PoW consensus mechanism.
\end{itemize}

Miners collect transactions in the mempool, arrange them in a block, and solve the PoW puzzle. Miners get to keep the transaction fees offered by transaction senders. Often, miners keep their transaction mempool sorted and add transactions in their block in descending order of transaction fees. We see how this could be leveraged by greedy parties in Section~\ref{ssec:PGA} and what challenges it leads to.

\subsubsubsection{Proof-of-Stake}
In PoS-based blockchains, participants interested in publishing their block must stake some of their blockchain's native token. While their tokens are staked, they can't be used in transactions. Though in all PoS protocols, a participant's probability of publishing a block is proportional to its fraction of the total stake, different PoS mechanisms propose their own randomization methods to select which \emph{staker} gets to publish the block. We discuss Ethereum's PoS briefly in the following section.

\subsubsubsection{Proposers, Validators, and Committees in Ethereum}
A staker can create multiple validator clients, each of which stakes 16 to 32 ETH. Blocks are published in slots (some of which can be empty) of $12$ seconds. $32$ slots make one epoch. Each slot is assigned a \emph{proposer} and a committee of validators. The proposer is responsible for publishing the block for their assigned slot. Committee members attest to the blocks from the proposer if they consider them valid.

\subsubsection{Incentives}
In PoW, the block consists of a \emph{coinbase} transaction in which the miner claims a mining reward. After the fourth halving on 19th April 2024, the block reward for Bitcoin was reduced to $3.125$ BTC. If a block mined by a miner is not accepted, the miner doesn't receive the reward.

In Ethereum, validators get rewards for attesting blocks that most validators attest to. They are penalized for failing to attest or attesting blocks that aren't finalized. Proposers of the block get larger rewards for publishing an accepted block. Participants can get slashed (i.e., penalized by losing some staked ETH) for malicious behavior like double-voting by attestors or double proposal by the proposer. Slashing rewards are received by the proposer who reports the malicious behavior.

\subsection{Smart Contract}
\label{ssec:SC}
A transaction in Bitcoin is a small program in scripting language \texttt{Script}. Using \texttt{Script}, one can write a transaction as a contract. Limited capabilities of \texttt{Script} limit the number of possible attacks. Using \texttt{Script}, one can execute different types of contracts on Bitcoin. Vitalik questioned, \say{Why restrict contracts to only monetary transactions? Why not anything that has value?} It leads to the realization of Nick Szabo's concept of \emph{smart contracts}~\cite{Szabo2018SmartCB}. Towards this, Vitalik introduced \emph{Ethereum}.

Ethereum and some similar blockchains support a Turing-complete instruction set and allow a wider variety of contracts to be written and executed on the blockchain. Smart contracts are programs that reside on blockchain and have their state (balance and variables) and functions. We discuss Ethereum's smart contract specifics in the next section.

\subsubsection{Ethereum}
Ethereum as a blockchain is a virtual machine known as \emph{Ethereum Virtual Machine} (\evm). The state of the machine evolves with each transaction, which is a computational payload. These computations can be arbitrary, and the transaction sender needs to pay gas fees to the block publisher to run their transactions. The native currency of Ethereum is \emph{Ether}, and its ticker is \emph{ETH}. Ethereum is arguably the most popular chain that allows smart contracts and has the most DeFi activity. To improve scalability, many chains derive security from Ethereum by committing their states to Ethereum, such as Polygon, StarkNet, Arbitrum, etc. Such blockchains are called layer 2 or \emph{L2}, while Ethereum is considered layer 1 or \emph{L1}.


\subsubsubsection{\evm}
As mentioned above, the Ethereum blockchain can be imagined as a single computer where users can request arbitrary computations by sending transaction requests in the mempool. Transactions list opcodes/instructions to be executed. An Ethereum address originates the transaction and is signed by the private key corresponding to the address.
 
The \evm\ specification is provided by the Ethereum Yellowpaper~\cite{ethereumyellowpaper}. Ethereum execution clients follow this paper's specifications.

\subsubsubsection{Gas}
As \evm\ is Turing complete, users can request arbitrary computations. A block in Ethereum needs to be finalized every $12$ seconds; thus, computations on \evm\ must be limited. The protocol ensures this as follows. The computation required by each opcode is represented in units of \emph{gas}. Users need to pay a base fee and optionally a priority fee for their transactions, calculated as:
$$Total\_fee = gas\_units \times (base\_fee + priority\_fee)$$

If the gas fee offered by the user falls short of the total gas fee required (more appropriately, if the gas is exhausted despite computations being left), the block creator can stop the execution and include the transaction as a failed transaction in the block while charging the gas fee offered. We now briefly explain how DeFi is executed on \evm.
\subsection{Decentralized Finance}
\label{ssec:DeFi}

With fungible and non-fungible tokens possessing monetary values, ownership transfers, and exchanges are required. Smart contracts allow these to be completely decentralized, removing trust assumptions required in traditional finance. DeFi is the ecosystem that caters to all finances happening on-chain, governed by smart contract logic, which is auditable by everyone.

Some decentralized exchanges (DEX) maintain an order book. There can be massive delays as these require two trading parties to have complementary requirements available simultaneously. \emph{Automated market makers} (AMMs) bypass this requirement. We discuss AMMs in the next section.

\subsubsection{Automated Market Makers}
The most commonly used DeFi application is swapping one token for another. AMMs allow users to swap two tokens (most commonly, native currency like Ether and one other token) without requiring another party. There are various ways of defining the swapping mechanism for an AMM. \emph{Constant function market makers} (CFMM) achieve this by maintaining a pool of the two tokens such that the product of their supply remains constant.
$$x \times y = k$$
where $x$ is the token$_1$'s supply, $y$ is token$_2$'s supply and $k$ is known as the \emph{invariant}.

Let the current supply of token$_1$ be $x_0$, and the current supply of token$_2$ be $y_0$. If a user wants to swap $a$ token$_1$ and receive some token$_2$ (= $b$), the new supplies will be $(x_0 + a)$ for token$_1$ and $(y_0 - b)$ for token$_2$. As their product remains equal to the invariant $k$, the amount of token$_2$ the user gets is:
$$b = y_0 - \frac{k}{(x_0 + a)}$$
We ignore fees for the above example. Uniswap V1 whitepaper~\cite{uniswapV1} delves into more details.

\subsubsection{\mev\ in Decentralized Markets}
The curve $x \times y=k$ is a hyperbola symmetric around the line $y=x$. It can be shown that as the supply of token$_1$ increases, diminishing amounts of token$_2$ are received. If $x \xrightarrow{} \infty$, then $y \xrightarrow{} 0$. For such $x$ and $y$, any added supply $\Delta x$ will give negligible token$_2$, but if for any finite $\Delta y$, huge amounts of token$_1$ will be swapped. The unreasonable price discourages parties from swapping. If the latter is the case, the parties get the system closer to equilibrium. Earning rewards to get the system back to equilibrium is called \emph{arbitrage}. Arbitrage is considered essential for the smooth functioning of markets.

Block creators can bring the system back to equilibrium by executing an inverse swap after every AMM swap by users. Thus, arbitrage is also considered a type of \mev. Suppose the block publisher doesn't have enough tokens to perform arbitrage. In that case, they can buy tokens before the user's transaction by placing them before the user's transaction and then performing arbitrage after the user's transactions, placing the block publisher's transaction before the user's transaction is called \emph{frontrunning}, and placing it after the user's transaction is called \emph{back-running}. Doing both together is called \emph{sandwiching}. All these are also types of \mev. We discuss examples of each in the next section.


\subsubsection{Example \mev\ Extraction}

\subsubsubsection{Frontrunning}
Frontrunning involves a miner or trader intercepting and prioritizing their transaction in front of another user's transaction to profit from the anticipated price movement caused by that transaction.

Suppose trader $T_i$ decides to buy a large amount of token$_x$ on DEX$_1$.  These will likely cause a temporary increase in the price of token$_x$ on DEX$_1$ due to the sudden decrease in supply. $T_i$ places a large market buy order on DEX$_1$. Trader $T_j$, who monitors pending transactions, anticipates this price rise and aims to profit by executing their transaction to buy token$_x$ just before $T_i$'s buy order at a lower price. $T_j$ submits his own transaction to buy token$_x$ on DEX$_1$ with higher gas fees than $T_i$'s transaction. Miners typically prioritize transactions based on gas fees, and as $T_j$'s gas fee is higher than $T_i$'s transaction, $T_j$'s transaction is included before $T_i$'s transaction. Suppose $T_j$'s transaction is successfully included in a block before $T_i$'s. In that case, it will affect the price of token$_x$ before $T_i$'s transaction is executed, allowing $T_j$ to front-run $T_i$ and potentially profit from the price movement.

\subsubsubsection{Backrunning}
Backrunning is the opposite of frontrunning, where the attacker tries to anticipate and exploit a price decrease instead of an increase. The attacker aims to submit their transaction just after the target transaction but before it gets included in the block.

Suppose trader $T_i$ decides to sell a large amount of token$_x$ on  DEX$_1$.  These will likely cause a temporary decrease in its price on DEX$_1$ due to the sudden increase in supply. $T_i$ places a large market sell order on DEX$_1$. Trader $T_j$, who monitors pending transactions, anticipates this price drop and aims to profit by executing their transaction to buy token$_x$ just after $T_i$'s sell order. $T_i$ submits his own transaction to buy token$_x$ on DEX$_1$ with slightly lower gas fees than $T_i$'s transaction. Miners typically prioritize transactions based on gas fees, and as $T_j$'s gas fee is only slightly lower than $T_i$'s transaction, $T_j$'s transaction is included right after $T_i$'s transaction. If $T_j$'s transaction is successfully included in a block right after $T_i$'s, it can sell the token$_x$ it bought at the higher price after the rebound, earning a profit from the price discrepancy.

\subsubsubsection{Sandwiching}
Sandwiching involves placing two transactions around a target transaction, effectively ``sandwiching" it between them. The attacker places a transaction before and after the target transaction to capture value from both sides of the trade.

Suppose two decentralized exchanges, DEX$_1$ and DEX$_2$, are trading the same token, token$_x$, at prices $A$ and $B$ such that $A < B$. Trader $T_i$ decides to execute the arbitrage trade by buying token$_x$ on DEX$_1$ and selling it on DEX$_2$. $T_i$ submits its buy order on DEX$_1$, which will land in the mempool pending and waiting to be confirmed. Trader $T_j$ monitors pending transactions, observes $T_i$'s pending transactions and submits its buy order on DEX$_1$ with a sufficiently high gas fee and sell order transaction with a slightly lower gas fee than $T_i$'s buy order. Miners include $T_j$'s buy transaction before $T_i$'s buy order and its sell transaction right after. Suppose $T_j$'s transactions are successfully included as per this order. In that case, it will affect the price of token$_x$ before $T_i$'s transactions are executed, allowing $T_j$ to front-run and then back-run $T_i$'s transaction and potentially snatching the arbitrage opportunity.


\section{\mev\ Landscape in Ethereum 1.0}
\label{sec:mev_landscape_1}
In summary, multiple avenues exist to extract profit by reordering the transactions. Such value extracted is referred to as \mev. The way \mev\ is materialized has evolved rapidly. This report aims to provide a summary of the evolution.

Ethereum has an account-based model in which each account has a balance. Each account and account balance is stored globally in the Ethereum Virtual Machine. Ethereum started as a PoW-based blockchain before shifting to PoS, where producing a new block required miners to select transactions from the mempool and compete to solve a complex puzzle. The first miner to solve got to add the block. Miners were rewarded with a block reward for successfully mining a block and received transaction fees from the included transactions. Miners were in complete control of the block's inclusion, exclusion, and ordering of transactions, allowing them to frontrun/backrun/etc. Thus, \mev\ was initially referred to as \emph{Miner Extractable Value}.

\subsection{Priority Gas Auctions}
\label{ssec:PGA}

\subsubsection{Introduction}
Users, i.e. transaction creators, express their intention to pay for the transaction inclusion and order preference in the block using the transaction fee. As rational players, miners are incentivized to select the transaction with the highest fee to include in the block. As \mev\ opportunity is limited and episodic, the competition to claim the opportunity is high. On Ethereum, this competition is observed mainly through front-running and back-running. When an \mev\ transaction (say, an arbitrage transaction) is seen in a public mempool, searchers (looking for \mev\ opportunities) send similar transactions to capture the same arbitrage with an increased transaction fee in the form of gas, as transactions with a higher fee are ordered above transactions with a relatively lower fee. As more searchers see such transactions within the same slot, they bid to the miner with higher gas. This keeps happening till the end of the slot, resulting in an all-pay English auction, also known as \emph{Priority Gas Auctions} (PGA). 

\begin{figure}[!h]
    \centering
    \includegraphics[width=\linewidth]{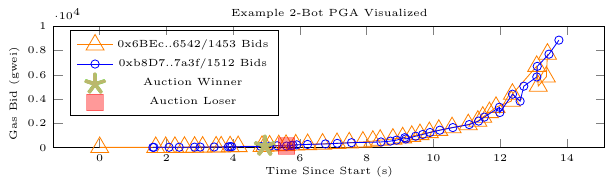} \\
    \scalebox{0.65}{
          \hspace{-6mm}
         
\newcommand{\specialcell}[2][c]{%
  \begin{tabular}[#1]{@{}l@{}}#2\end{tabular}}

\begin{tabular}{c|c|l|c|l}

\specialcell{\textbf{Seconds}\\ \textbf{Elapsed}} & \textbf{Quantity @ Price Bid} & \hfil\textbf{Ethereum Transaction Origin (Public Key Hash)} & \textbf{Nonce} & \hfil\textbf{Transaction Hash} \vspace{1mm}\\

0.000 & 192085 @ 25.10 & \color{orange} 0x6BEcAb24Ed88Ec13D0A18f20e7dC5E4d5b146542 \color{black} & 1453 & 0xd32653ca9694a6d1299335f3c04f74cc159bee48c1d32d3a421db08c638ffc78 \\
1.593 & 231520 @ 25.00 & \color{blue} 0xb8D76f4BC2518F8eb508bf0Ccca76f8F9DD57a3f \color{black} & 1512 & 0xb901e6dc2c229fd9105448fcc23eaebdedb476c21b6c6e7ddf8d2df4e838d2c7 \\
1.624 & 231520 @ 28.75 & \color{blue} 0xb8D76f4BC2518F8eb508bf0Ccca76f8F9DD57a3f \color{black} & 1512 & 0x9f592504eb71a7452b7a395a7f5ecd34eaa5d090da1162e74221562af54c8f67 \\
1.679 & 227534 @ 28.81 & \color{orange} 0x6BEcAb24Ed88Ec13D0A18f20e7dC5E4d5b146542 \color{black} & 1453 & 0x83e2a6774654a9540c3fad8837afcc88b4c932ab2374819254f887305c3a4b22 \\

... & ... & ... & ... & ... \\

4.949 & 227534 @ 134.02 & \color{orange} 0x6BEcAb24Ed88Ec13D0A18f20e7dC5E4d5b146542 \color{black} & 1453 & \color{olive} 0xc889bd13594f75e4dd824f04f0c2ad03896cb7ec6518df02455e9560367bb9c4 \color{black} \\

5.599 & 231520 @ 133.76 &  \color{blue} 0xb8D76f4BC2518F8eb508bf0Ccca76f8F9DD57a3f \color{black} & 1512 & \color{red} 0xaa86d782328c0c9c422e3f2a3170ff41ae21a27ad395c48db76b0080898f85db \color{black} \\

... & ... & ... & ... & ... \\

13.383 & 227534 @ 5834.77 & \color{orange} 0x6BEcAb24Ed88Ec13D0A18f20e7dC5E4d5b146542 \color{black} & 1453 & 0xb0dc97140394c5f65332ebc459d5e66f89099dbb4d335c866b32280270102858 \\
13.416 & 227534 @ 7716.48 & \color{orange} 0x6BEcAb24Ed88Ec13D0A18f20e7dC5E4d5b146542 \color{black} & 1453 & 0x1825be6951577e72a1dafc8de564ce1ccfe5d284173e11e77b2e7f6b1b44571c \\
13.462 & 231520 @ 7701.08 & \color{blue} 0xb8D76f4BC2518F8eb508bf0Ccca76f8F9DD57a3f \color{black} & 1512 & 0xa9823358c99149f0e6343c604c35988468d01d02868437d8251b3cee282dc92b \\
m13.759 & 231520 @ 8856.24 & \color{blue} 0xb8D76f4BC2518F8eb508bf0Ccca76f8F9DD57a3f \color{black} & 1512 & 0x366c30a534b5f3d8a6d251f97d401997624d1fe8d3af07ede4d19105dc970942 \\
    \end{tabular}

     }
    \caption{Observed PGA on Ethereum \cite{DBLP:journals/corr/abs-1904-05234}}
    \label{fig:exampleauction}
\end{figure}   

As nodes try to outbid others, the transaction fee submitted by nodes rises drastically. 
Consider the example of observed PGA mentioned in \cite{DBLP:journals/corr/abs-1904-05234}. Two bots submitted 85 bids within 14 seconds, as shown in Figure \ref{fig:exampleauction}. The graph shows the gas bids of two observed bots, while the bottom table details the first and last two bids placed by each bot and the two mined bids (center). 

Two accounts compete with one another for priority: \href{https://etherscan.io/address/0x6BEcAb24Ed88Ec13D0A18f20e7dC5E4d5b146542}{0x6B...42} with nonce 1453 and \href{https://etherscan.io/address/0xb8D76f4BC2518F8eb508bf0Ccca76f8F9DD57a3f}{0xb8...3f} with nonce 1512. While the auction ended in 4.94 seconds, bot 0x6B...42 issued 42 transactions in 13.4 seconds, and bot 0xb8...3f issued 43 transactions in 12.1 seconds with increasing gas fees. Transaction 0xc8...c4, issued by bot 0x6B...42, is the one that is ultimately mined with priority; it is indicated by a green star. This bot, which pays the full gas fee to win, is regarded as the winner. The red square transaction, hashed 0xaa...85db by bot 0xb8...3f, is also mined and added to the final block as miners collect transaction fees from failed transactions.


\subsubsection{Tranasaction Flow}
\begin{figure}[!ht]
    \centering
    \includegraphics[scale=1]{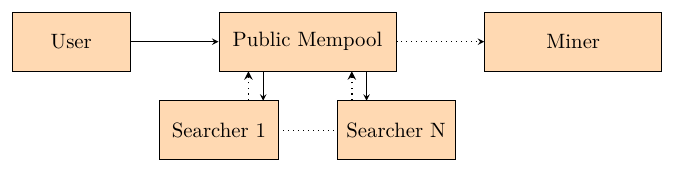}
    \caption{Transaction flow in Priority Gas Auctions}
    \label{fig:PGA Flow}
\end{figure}

The transaction flow in the case of PGA can be understood from Figure \ref{fig:PGA Flow}. A user sends an \mev\ transaction on the network, which is received by the public mempool of the miners and searchers. Searchers identifying the \mev\ transaction keep creating new transactions with a relatively higher fee than those already submitted in the mempool. Miner orders the transaction based on high to low transaction fees and publishes it on the network.

\subsubsection{Issues with Priority Gas Auctions}
Miners maximize their profits by including the most profitable transactions, sorting the transactions in decreasing order of the transaction fee, and making the best block. While this was profitable to miners, there were many negative externalities.

Firstly, the network gets congested with similar transactions trying to capture the \mev\, bloating the mempool. Secondly, the miner is incentivized to pick up the transaction with a higher fee; therefore, low-fee transactions that don't involve \mev\ might face stagnation. While low-fee transactions usually experience more delays than high-fee transactions, the \mev\ auctions incur more delays for non-\mev\ transactions. Thirdly, as these auctions are on-chain, many \mev\ transactions are included on the final block (due to high gas fees). However, the first transaction captures the \mev, and the rest of the transactions trying to capture it revert, resulting in a wastage of blockspace. Lastly, since miners can get the transaction fee even from transactions that revert, miners are motivated to include such transactions, resulting in blockspace even after the \mev\ is captured. To avoid mempool bloating and network congestion, Flashbots \cite{flashbots} have proposed an off-chain auction, which we refer to as \emph{Bundle Auctions}.

\subsection{Bundle Auctions}
\label{ssec:bundle_auctions}

\subsubsection{Introduction}
PGA has serious problems mainly since the auctions were online and involved a public mempool, and therefore, there is a need to shift the auctions to off-chain. Flashbots \cite{flashbots} is a research and development organization formed to mitigate the negative externalities of \mev. In January 2021, Flashbots launched the \mevg\ to improve the efficiency of the Ethereum network. \mevg\ is an upgraded fork of the Ethereum client \emph{Geth}, and it introduced the concept of bundles. A bundle refers to a collection of transactions packaged together and submitted to miners or validators for inclusion in a block. \mevg\ modified Geth's transaction mempool to include a field $mevBundles$, which stores a list of \mev\ bundles and enables sealed bid auctions for bundle inclusion \cite{mevgeth}. It provides an API for miners and a relay service, which receives the searcher's bundles and forwards them to miners. Miners include the most profitable bundles in the block proposed.

\subsubsection{Transaction Flow}
\begin{figure}[!ht]
    \centering
    \includegraphics[width=0.8\columnwidth]{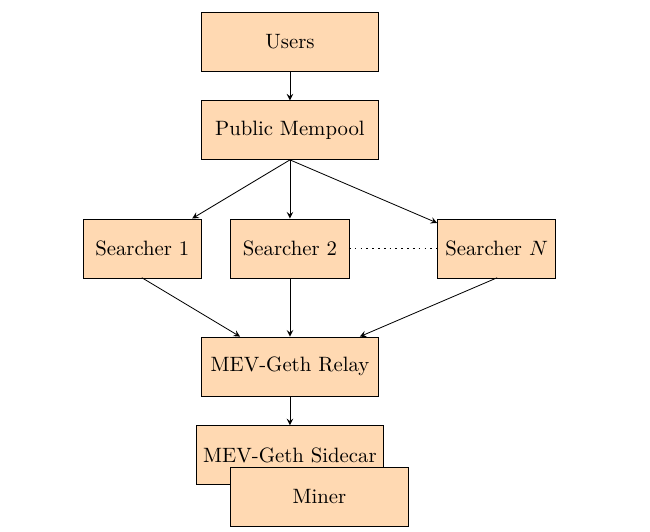}
    \caption{Transaction Flow in Bundle Auctions}
    \label{fig:Bundle Flow}
\end{figure}

The transaction flow in the case of Bundle Auctions can be understood from Figure \ref{fig:Bundle Flow}. Users submit transactions to the blockchain network, which lands in the public mempool. Multiple searchers who identify the \mev\ opportunity compete in a sealed-bid auction with a miner for their transaction inclusion via the \mevg\ relay. Miner receives the bundles on its \mevg\ sidecar, selects the most profitable bundles and transactions from mempool, and produces the block. 

\subsubsection{Issues with Bundle Auctions}
While \mevg\ addresses the bloated mempool and wasted blockspace, it requires searchers to trust other participants. A miner could still choose to censor a bundle and create its own, or the relay might reorder or fail to forward the bundle.


\section{\mev\ Landscape in Ethereum 2.0}
\label{sec:mev_landscape_2}
As Ethereum shifted Proof-of-Stake, production of the next block is no longer decided by computationally intensive competition but rather by their stake. In each slot, a proposer is chosen randomly from the set of proposers, a node with a stake of at least 16 ETH on the network. Post-merge, block proposing, and building were separated, often known as the proposer-builder separation ($\texttt{PBS}$). Node proposers in the PoS paradigm are exclusively responsible for validating blocks and play no part in block building. The separation between proposers and builders fosters an open market where block proposers can acquire blocks from block builders. These builders compete with one another to construct the block, offering the highest fee to the proposer, which we refer to as \emph{Block Auctions}.

\begin{figure*}[!ht]
\includegraphics[scale=0.5]{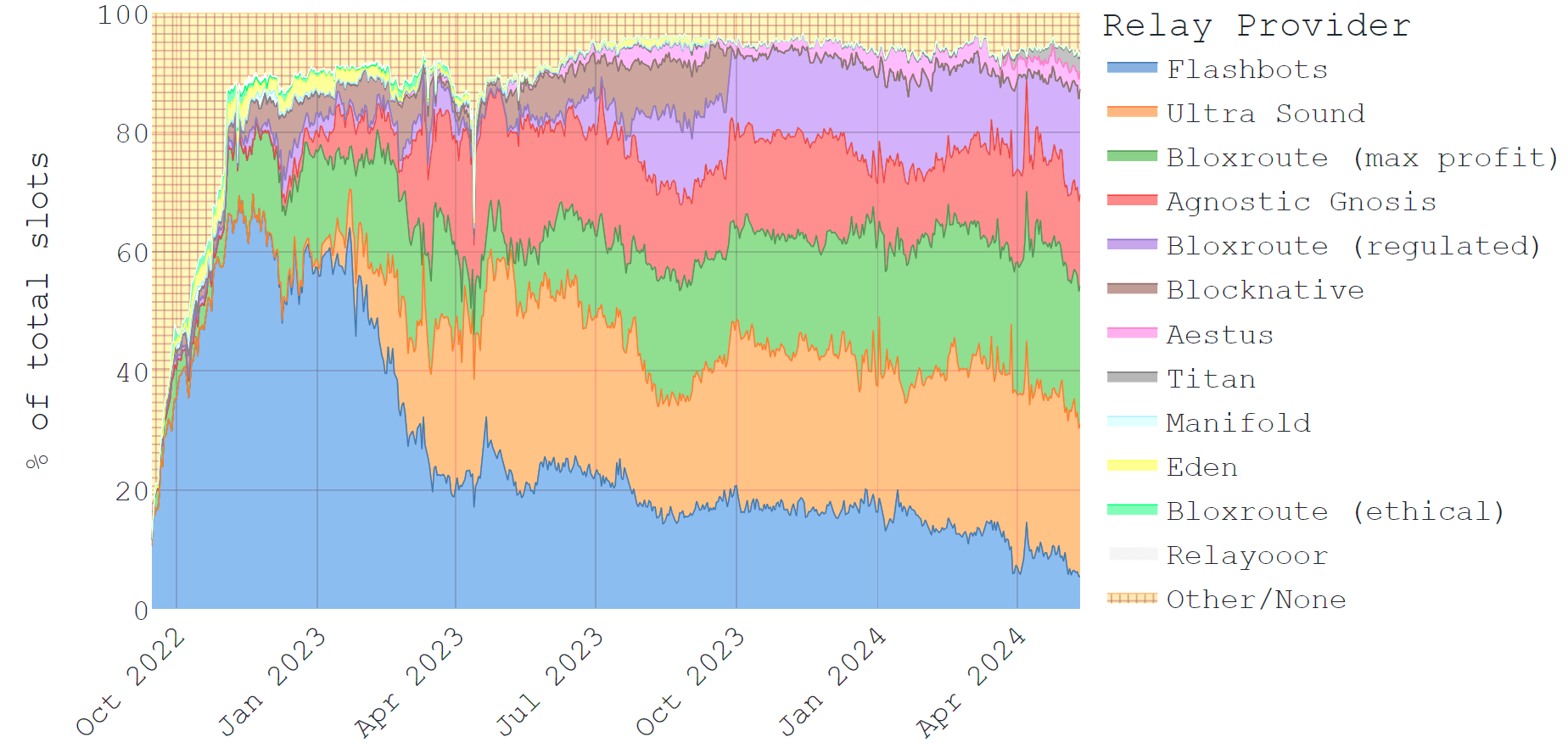}
\caption{Percentage of total slots dominated by various relays since the merge \cite{mevBoostDash} }
\label{fig:pslots}
\vspace{-4mm}
\end{figure*}

\subsection{Block Auctions}
\label{ssec:block_auctions}
\subsubsection{Introduction}
Adapting to the new paradigm, Flashbots have introduced \mevb, an open-source software client for proposers similar to \mevg\ used by proposers, and its relay as an intermediary aggregator between builders and proposers. Relays collect the most profitable blocks from block builders and relay the headers and bids to the block proposer. Builders concentrate exclusively on collecting transactions and crafting efficiently optimized blocks. These blocks are then relayed to proposers via a relay mechanism. Using \mevb, builders bid to the proposer via relay. Note that the relay has full access to the block. Flashbot's relay service quickly became popular, and various players have started offering similar services to builders, such as Bloxroute, Ultra Sound, Dreamboat, and Agnostic Gnosis. Figure \ref{fig:pslots} shows the percentage of total slots dominated by various relays since the merge.

Unlike with \mevg, a proposer cannot see the block, and only upon signing the block header does the relay release the block to the network. This prevents censorship by the proposer. Instead of submitting the bundles to the proposer, searchers now bid for inclusion in the block to the builders in the form of a transaction fee for builders. Builders must build the block with mutually exclusive bundles and thus select the most profitable bundle. The builder creates the most profitable block and bids in a sealed-bid first-price auction with the proposer using the \mevb\ relay, where the relay forwards the block headers and bids to the proposer. The proposer selects the most profitable block and responds with a signature on the respective header as a commitment back to the relay, after which the relay releases the block.

\begin{figure*}[!ht]
\includegraphics[scale=0.5]{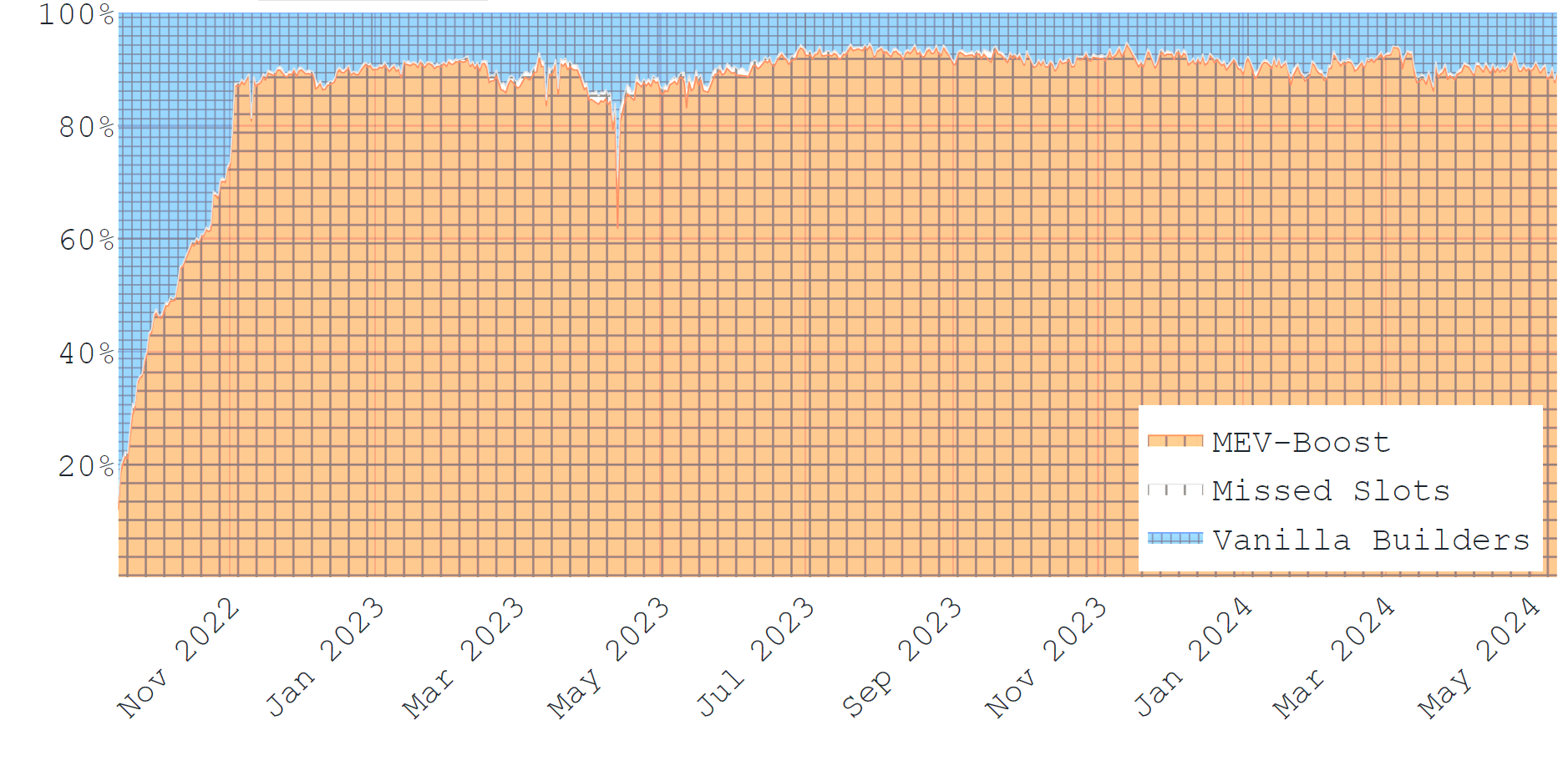}
\caption{ Percentage of total builders using \mevb\ against vanilla builders post-merge \cite{mevBoostDash} }
\label{fig:MEVBS}
\vspace{-4mm}
\end{figure*}

\begin{figure}
\centering
\includegraphics[width=0.7\textwidth]{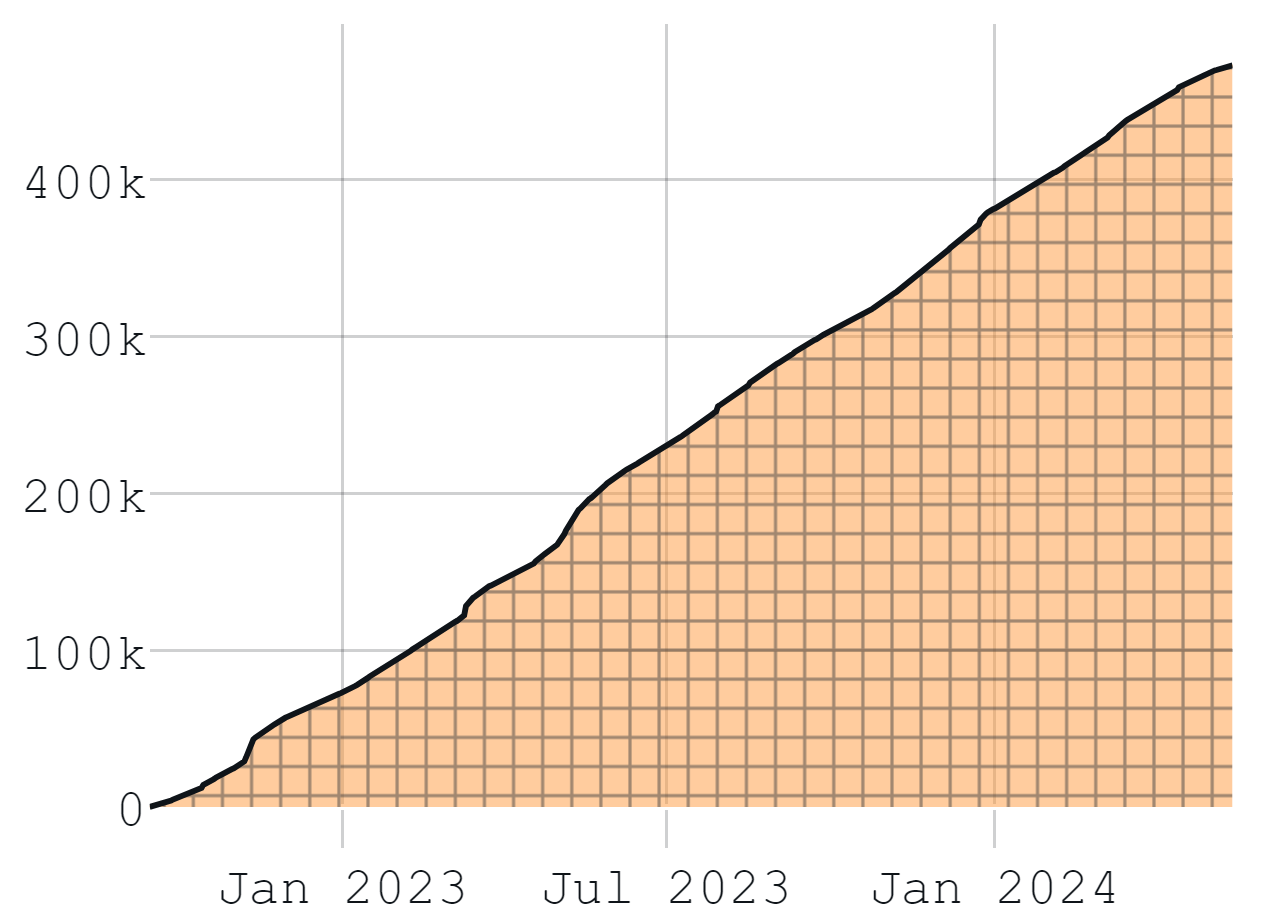}
\caption{Cummulative \mev\ generated in ETH through \mevb\ \cite{mevBoostDash} }
\label{fig:MEVBCGP}
\vspace{-4mm}
\end{figure}

Most builders adopted \mevb\ to submit the blocks to proposers. Figure \ref{fig:MEVBS} shows the percentage of blocks mined through \mevb\ over vanilla builders post-merge, and Figure \ref{fig:MEVBCGP} shows the total cumulative \mev\ generated using \mevb.

Further, the relay can use public escrows for data availability \cite{relayEscrow}. Escrows receive the full execution payloads from relays. proposers receive block headers from relays with an indication of the block value of each header. The proposer selects the most valuable header, signs it, and returns it to the relay and the escrow to propagate the respective block to the network. However, the proposer and relay must trust the escrow to save and propagate blocks. Before signing a block header, a proposer can query different escrows that it trusts to certify the availability and validity of the entire block and only sign if it gets at least one positive response. 

\subsubsection{Transaction Flow}
\begin{figure*}[!ht]
    \centering
    \includegraphics[scale=0.9]{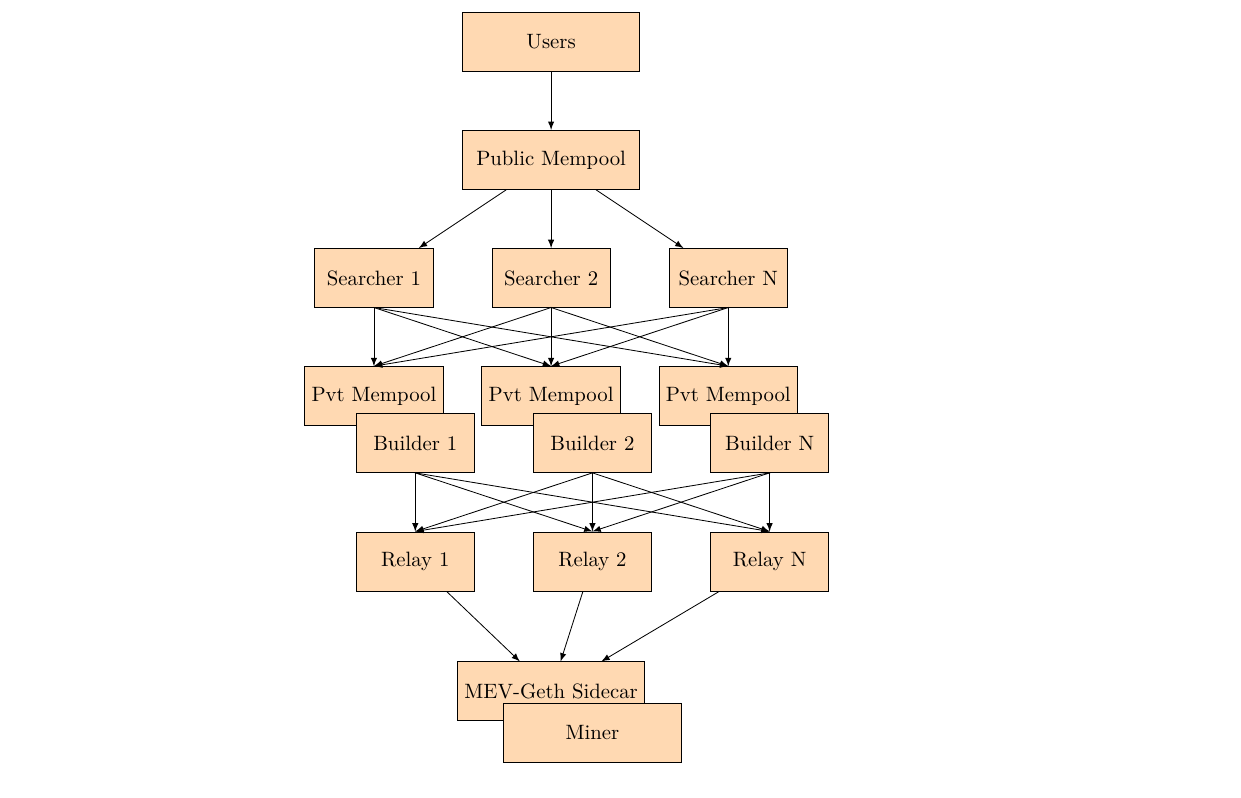}
    \caption{Transaction flow in Block Auctions}
    \label{fig:Block Auctions Flow}
\end{figure*}

Figure \ref{fig:Block Auctions Flow} shows the transaction flow in block auctions. Users submit transactions to the blockchain network, which lands in the public mempool.

Searchers who identify \mev\ transactions from the public mempool or any other profitable revenue opportunity bid for their bundle inclusion to block builders. Block builders create the most profitable block from searcher bundles and public mempool transactions and bid in PBS auction with the proposer via relays. The proposer selects the most profitable block and publishes its header, after which the relay releases the block to the proposer to publish.

\subsubsection{Issues with Block Auctions}\label{sssec: issues with block auctions}

 Although this solution has been widely adopted (with \mevb\ mining approximately 90\% of the blocks by 2022 \cite{ratednetwork}), there were still issues present within this architecture. Firstly, the dependence on third-party relays doesn't seem to have any incentive to perform honestly, which is concerning. Relays are trusted by builders for routing the block payload and trusted by proposers for block validity and data availability \cite{relayFund}. As reported by Flashbots \cite{postmortem}, a vulnerability in the ultrasound relay allowed a malicious proposer to steal its \mev\ worth \$20M. The relay did not check if the proposer signed a valid block header, and the relay released the block to the proposer. This allowed the proposer to extract transactions from the block and steal \mev. Another potential attack has been discovered in the same report where the proposer requests the block header from the relay when its slot is about to end and signs it at the end of the slot. The relay would release the block to the proposer, but the proposer missed the slot. This attack allows the proposer to grieve the relay for missing the block, causing the block to be revealed, though the proposer might not be able to capture that \mev. To mitigate this attack, relays adopted a cutoff timing in the proposer’s slot; after the cutoff time, the relays will no longer return a block to the proposer.

 While such vulnerabilities could be patched, the relays themselves can still choose to act maliciously and steal the \mev. This dependence on the relay could affect the liveness of transactions and the privacy of the builder's block, granting an unfair advantage to proposers or relays.
 
 Secondly, the block builder can still censor any transaction or bundle. However, this might affect its reputation over time, resulting in the searcher sending the bundle to other builders. 

 Most often, \mev\ extraction process is done by searchers and builders that derive value out of users' transactions. Searchers choose potential \mev\ extractable transactions and strategically insert their transactions with user transactions to create profitable bundles. The bundled transactions are further placed in the block and reordered by the builders. This extraction might adversely affect users' net transaction utility. Users preferred sending their transactions privately to builders to protect the transaction from being exploited.


\section{Private Order Flow}
\label{sec:private_order_flow}

\subsection{Introduction}
\label{ssec:introduction:private_order_flow}
As defined by Flashbots \cite{ofaFlash}, \emph{order} is defined as anything that changes the state of the blockchain. In the context of \mev, the notion of order is not just limited to transactions in the network but also includes intents from account abstractions. A collection of orders is called \emph{order flow}.

The \mev\ capture might adversely affect the transaction creators, aka users, who don't get anything from the \mev\ extraction. The users have to either wait long to get their transactions or, in the worst case, get adverse outcomes for their orders, such as worst execution prices due to front-running, where users might end up paying higher prices for their purchases or receiving lower prices for their sales. 

In the shift to private order flow, \emph{Order Flow Auctions} (OFAs) have emerged as a solution to help users compensate for the value their orders create, where users' transactions are auctioned to searchers. The auctioneer takes a small cut and shares the rest of the revenue with the users via the wallet service provider (if they are using one).  The first solution that resembled OFAs was KeeperDAO’s (later known as ROOK protocol) ``Hiding Game,” launched in February 2021 \cite{monoceros}. Users send their orders (intents/transactions) to a third party, which auctions off orders to searchers who wish to extract \mev\ from the exclusive order flow. A significant portion of the winning bids is sent back to the users.

There are two types of order flow - \emph{private} and \emph{exclusive}. Private order flow refers to orders sent directly to parties bypassing public mempool. For instance, \mevl\ (an RPC endpoint that allows transactions to be back-run and offers protection from front-run) sends order flows to multiple top-performing builders. Exclusive order flow refers to orders that are sent exclusively to one party. As highlighted in \cite{ofaIOSG}, more than 90\% of searcher flows of Symbolic Capital Partners 1 are exclusively sent to beaver-build builder, indicating vertical integration between them.

OFA is a mechanism that pays users the appropriate value for their order. Here, the dApps/wallets/custodians that users interact with are called \emph{Order Flow Originators} (OFO), and the auctioneer service is called \emph{Order Flow Providers} (OFP). Searchers or integrated builder-searchers are the bidders competing to get the exclusive rights to strategize and execute the orders.

\subsection{Transaction Flow}
\label{ssec:transaction_flow}
\begin{figure}
\centering
    \includegraphics[scale=0.75]{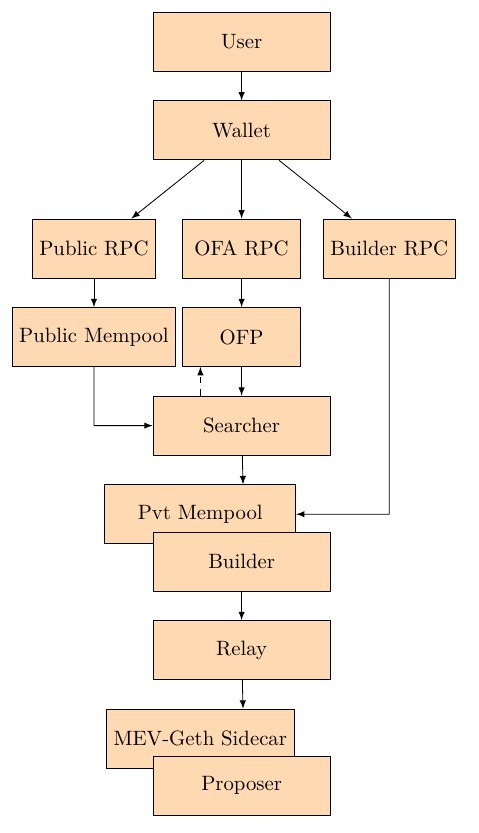}
    \caption{Private order flow stats for May 2024 \cite{mempoolpics}}
    \label{fig:PPOF}
\end{figure}

Figure \ref{fig:PPOF} illustrates the transaction flow with private order flow. The user sends the intent to its wallet  (either run locally or to a service provider), which creates a transaction and sends it to a public or private RPC endpoint based on the wallet configuration. The transaction lands in the public mempool if public RPC is used. On the other hand, if a private RPC like \mevl\ is used, the transaction is sent to the builder(s) private mempool, and if a private RPC of an OFP is used, the transaction is sent to the OFP. Searchers create profitable bundles from public mempool transactions and transactions from OFAs and send them to builders. A builder constructs the blocks from public mempool transactions, bundles from searchers, and those obtained directly from the private RPC. The builder bids in the PBS auction with the proposer via relay, and the proposer selects the most profitable block and publishes its header, after which the winning builder releases it.

\begin{figure}
    \centering
    \rotatebox{270}{\includegraphics[scale=1]{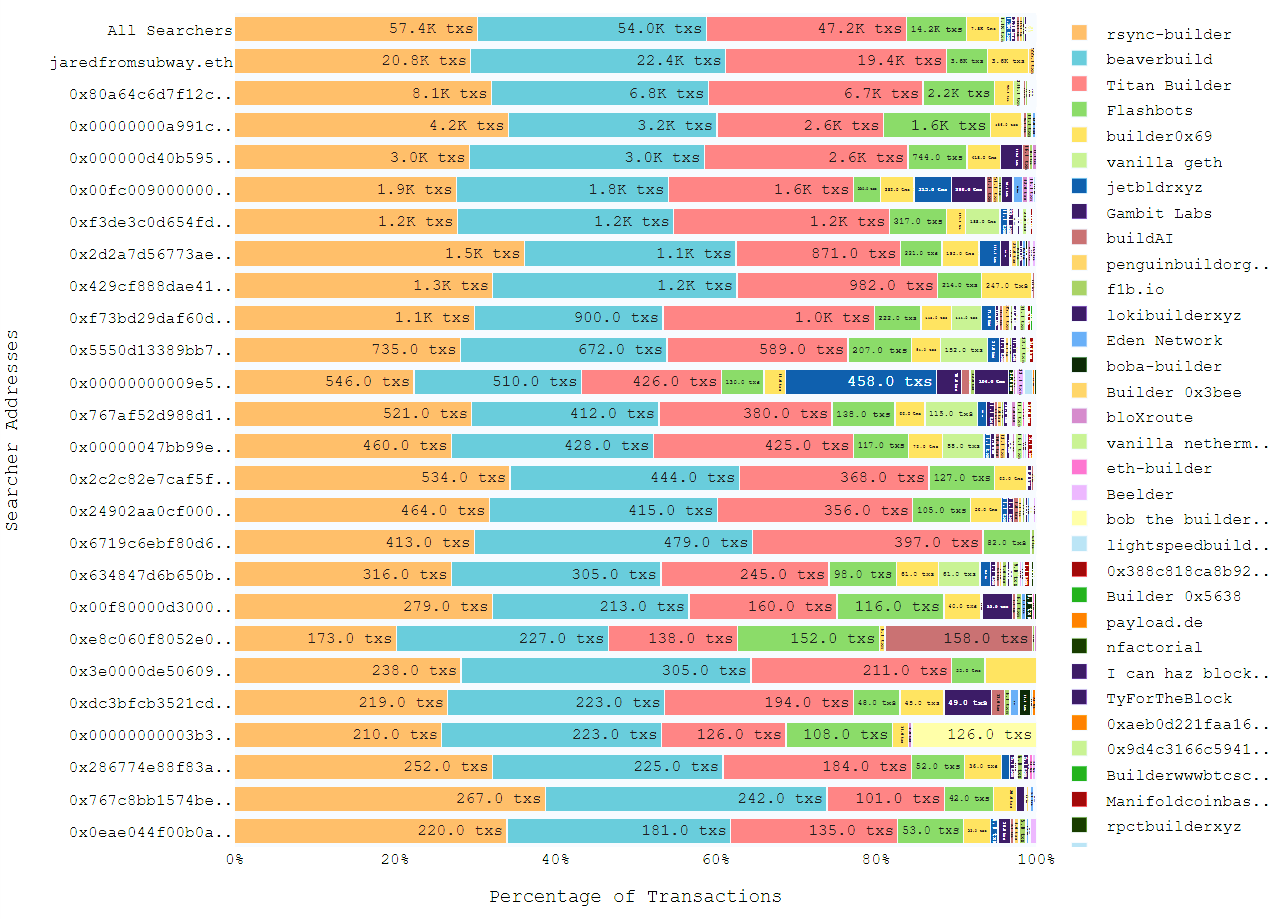}}
    \caption{Atomic searcher flows \cite{searcherbuilderpics}}
    \label{fig:bsd1}
\end{figure}

\begin{figure}
    \centering
    \rotatebox{-90}{\includegraphics[scale=1]{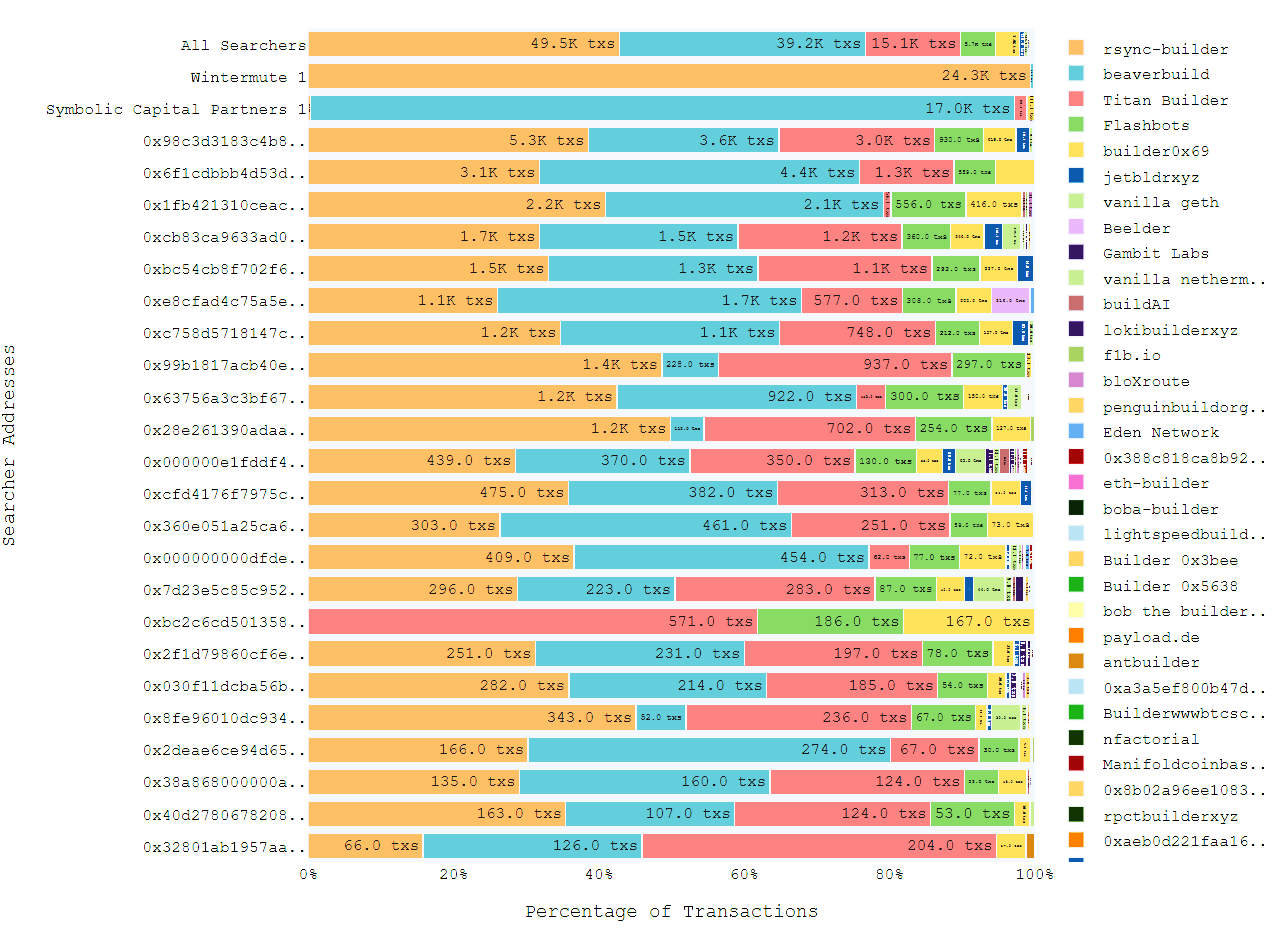}}
    \caption{Non-atomic searcher flows \cite{searcherbuilderpics}}
    \label{fig:bsd2}
\end{figure}
\subsection{Impact of Private Order Flow}
\label{ssec:effects_of_POF}

Builders are actively trying to compete for private order flows. Figure \ref{fig:PPOF} shows the percentage of private order flows and types of \mev\ extracted by various builders for May 2024. Under this competitive \mev\ market, builders began offering many lucrative services to gain more order flows \cite{ofaIOSG} such as:

\begin{itemize}
    \item \textbf{Transaction pre-confirmation}:
    Builder publicly agrees that if a user sends a transaction with greater than a certain priority fee, it will immediately send an (enforceable) message agreeing to include it and even send a post-state root to the user \cite{ofaSBC}. This is important for some use cases involving contention for the block space.
    
    \item \textbf{Front-running protection}:
    Builders promise not to front-run or opt to receive transactions from RPC endpoints that provide front-running protection, such as \mevl\ and Flasbots Protect. 

    \item \textbf{Revert protection}:
    If a certain bundle fails or reverts, the builder will not include it.
    
\end{itemize}

Private order flows can be divided into categories: searcher and user flows. Within searcher flows, we have atomic and non-atomic searcher flows. Atomic searcher flows include DEX-DEX arbitrage, sandwiching, and liquidation, and non-atomic searcher flow refers to CEX-DEX arbitrage. Figures \ref{fig:bsd1}, \ref{fig:bsd2} depict the percentage of transactions by various searchers to builders as of December 2023. The atomic searchers' flows are mostly sent to only the top three builders, possibly due to concern about the inclusion guarantee on a chain before \mev\ lost. Though it might seem in the searcher's best interest to send bundles to all the builders, according to Titan's research \cite{titanresearch}, it puts the searcher bundles at risk as the not-so-reputed builders might unbundle (may be due to poor implementation of block building), try to leak information due to long-tail strategies, gas-out-bidding by integrated builder-searcher builder trying to capture the same \mev. In the case of non-atomic searcher flows, we see a similar pattern but with few searchers often sending bundles exclusive to a specific builder. This highlights that there are vertical integrations between the searcher and the builder. Such entities are referred to as integrated builder-searchers.

\begin{figure*}[ht]
    \centering
    \includegraphics[width=\textwidth]{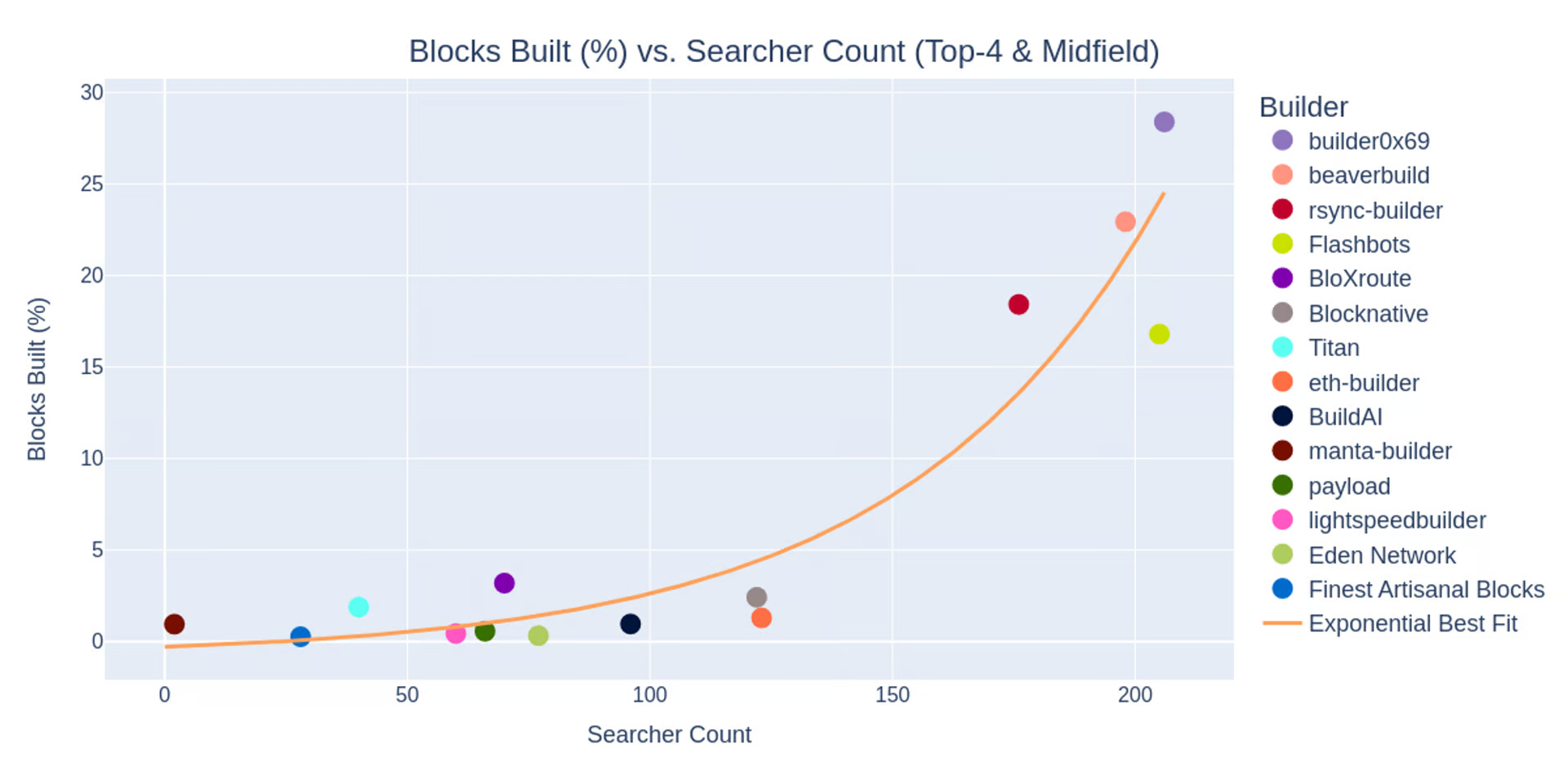}
    \caption{Builder and searcher dependence \cite{titanresearch}}
    \label{fig:bsd}
\end{figure*}

Figure \ref{fig:bsd} highlights the distinction between the top four builders and other mid-field builders based on the number of searchers they are connected to. The graph shows a positive correlation between the percentage of blocks produced and the number of searchers connected to the builders. While the graph shows a positive correlation, it is worth noting that the builder market share depends on various other factors, such as reputation and offered features. Among mid-field builders, there are builders with relatively more connected searchers yet underperform, such as Flashbots. This might be due to small infrastructure and inefficient strategies. On the contrary, builders like BloxRoute perform relatively better than others. This might be due to having its private relay reducing latency, large infrastructure, and other services such as Backrun service (BackRunMe) that give access to exclusive order flows.

When it comes to users, most often, users didn't control the \mev\ they create until OFAs came into existence as they send their orders to wallets or dApps, which are later exposed to public/private mempools or seachers. 

\begin{figure*}
    \centering
    \includegraphics[width=\textwidth]{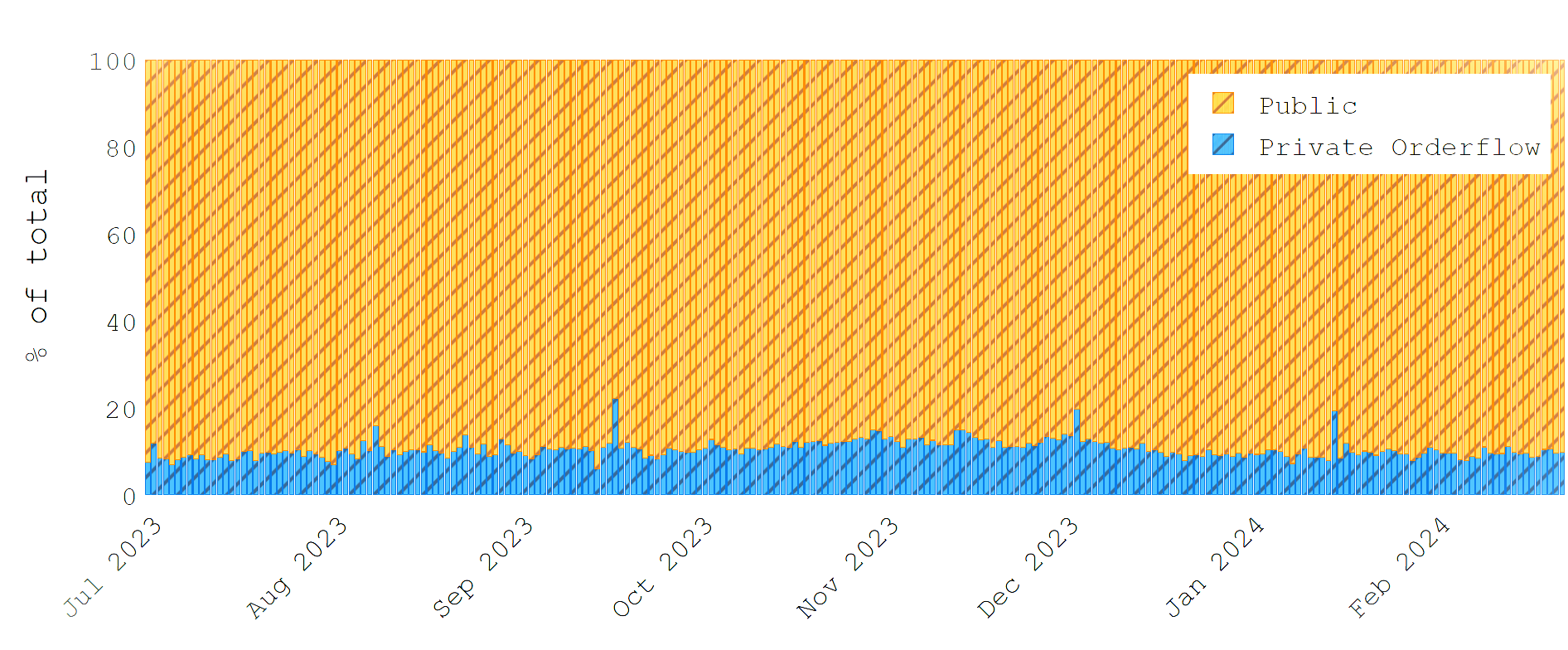}
    \caption{Percentage of private order flows from July 2023 to February 2024 \cite{mempoolpics}}
    \label{fig:PPOFT}
\end{figure*}

Figure \ref{fig:PPOFT} shows the percentage of public order flows in the network from July 2023 to Feb 2024. There is 12-13\% private order flow present on average throughout the period.

\subsection{Issues with Private and Exclusive Order Flows}
\label{ssec:issues_with_EOF}

\begin{figure*}
    \centering
\includegraphics[width=\textwidth]{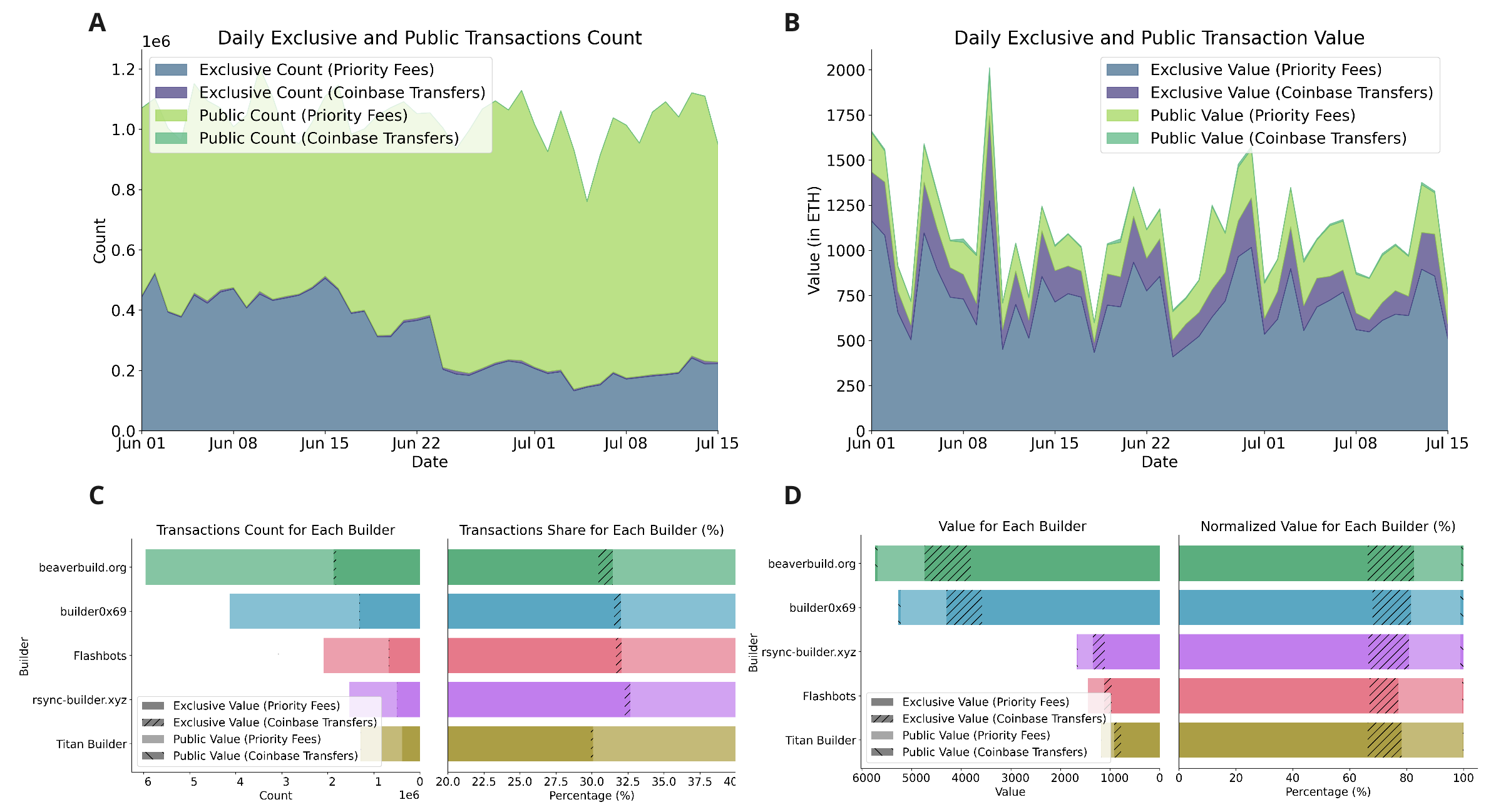}
    \caption{EOFs transaction count and value  \cite{BBPS}}
    \label{fig:EOF}
\end{figure*}

Block builders need to collect high-value order flow to create profitable blocks to beat competing builders; however, only a few get access to them. Consider Figure \ref{fig:EOF} that quantifies the number of top builders' public order flow and exclusive order flow and value transfer. The number of public order flows is large, and the number of exclusive order flows is less than 20\%. It can be seen that the value of exclusive order flow is close to 80\%. It signifies the value of exclusive order flow.

The block builder most likely to have their blocks chosen by the proposer is the one that consistently offers the highest value blocks. The builder that consistently presented the highest number of transaction bundles becomes the preferred destination for most OFOs submitting their bundles. Such a builder gets exclusive access to high-quality order flow and produces more profitable blocks, which can provide more services to OFOs. Such builders achieve better inclusion rates. Thus, the same builder becomes the highest-value block creator. Determining which block builders win bids, even in real-time, is relatively straightforward for searchers. This simple network effect sets in motion a cycle that could potentially result in centralization among builders where OFOs send more flow to the builder, providing the best UX and a high guarantee of block publishing.

This is seen empirically (refer Figure \ref{fig:bsd1}, \ref{fig:bsd2}), where searchers send bundles to only top builders. Also, 88.5\% of searchers send bundles to four or more block builders \cite{titanresearch}. This is because there is more than one competitive block builder, so sending to the top four companies has a higher chance of being captured. However, searchers need to trust that block builders will not steal \mev, so there is a tradeoff between the number of builders bundles are sent to and the cost of trust. Thus, transactions are concentrated among a few trusted, good-performing builders. 

Direct outreach is challenging in permissionless blockchains, and trust is hard to establish for new builders not tied to a public organization; block subsidization is often their only strategy to gain market share. Builder-searcher entities have an edge, as they do not always need external flow to win blocks. They can also push their searcher profits downstream to the builder, increasing their chances of submitting a winning bid. However, this might discourage other searchers from interacting with them due to potential concerns over front-running similar bundles. Being a high-economies-of-scale activity, the barrier to entry is also high, centralizing the market even further. In a nutshell: \say{\mev\ extraction is naturally a highly high-economies-of-scale and centralization-prone activity.}\cite{ecOfScale}

While OFAs address the problem of exclusively orderflow somewhat through auctions, two main challenges of designing an OFA auction are avoiding a trusted auctioneer and achieving low-latency guarantees and high-value capture.


\section{Current Developments}
\label{sec:current_development}
Some of the current proposals that the Ethereum community has put forth to develop the Ethereum 2.0 landscape are \emph{Enshrined \pbs} (aka e-\pbs) and \suave.
The e-\pbs\ proposals aim to remove the dependence on relayers from the transaction flow (which takes care of the issues discussed in Section \ref{sssec: issues with block auctions}.)
\suave\ aims to make the flow of transactions transparent. Consequently, the design disincentivizes builders from attempting bundle theft from searchers.
\subsection{Enshrined Proposer-Builder Separation}
\label{ssec:ePBS}
\say{Enshrined \pbs\ (e-\pbs) advocates for implementing \pbs\ into the consensus layer of the Ethereum protocol} \cite{epbs}, \say{Because there was no in-protocol solution at the time of the merge, Flashbots built \mevb, which became a massively adopted out-of-protocol solution for PBS that accounts for $\approx$90\% of Ethereum blocks produced.}

The following are the desired properties of an e-\pbs\ Market Design \cite{idealPBS}:
\begin{itemize}
    \item \emph{Untrusted proposer friendliness:} The Proposer is not required to be a trusted party in the protocol.
    \item \emph{Untrusted builder friendliness:} The builder is not required to be a trusted party in the protocol.
    \item \emph{Weak proposer friendliness:} Proposers need not have high bandwidth or technical sophistication.
    \item \emph{Bundle un-stealability:} Proposers should not be able to steal \mev\ from builders.
    \item \emph{Consensus-layer simplicity and safety:} Consensus layer (forking rules) should operate as before.
\end{itemize}



\subsubsection{2-Slot Proposer-Builder Separation}

\label{ssec:2_slot_PBS}
With emphasis on the above properties, Vitalik proposed \emph{2-Slot Proposer-Builder Separation}. The proposer publishes the \emph{exec body} in two consecutive slots in this design. Exec body refers to the contents of the block as created by the block creator (in this case, the proposer). \emph{Exec header} refers to the (intermediate) block header. The sequence of events in a slot pair is shown in Table \ref{tab:2slot events}, where $t$ indicates time.
\begin{table}[ht]
    \centering
    \begin{tabular}{l|l|l}
     $t$ (time in seconds) & Event & Explanation \\
     \hline
       $t\approx0^{-}$  & exec header publishing & Anyone can publish an exec header, which \\
        &   &   contains an exec block hash, a bid, and \\
        &   &   a signature from the builder. \\
       $t=0$  & beacon block deadline & Beacon block must include the winning \\
        &   &   exec header. \\
       $t \in [0,2.67]$  & attestations on beacon block & Only one committee attests to the \\
        &   &   beacon block. \\
       $t\in [2.67,8]$   & intermediate block deadline & The winning builder publishes an \\
        &   &   intermediate block, consisting of the \\
        &   &   exec block body and as many \\
        &   &   attestations on the beacon block \\
        &   &   as they can find. \\
       $t\in [8,10.67]$  & attestations on the intermediate block & The remaining $N-1$ committees attest \\
        &   &   to the intermediate block. \\
       $t\in [10.67,13.33]$   & aggregation of intermediate block & The proposer aggregates \\
        &  attestations &   the attestations to the intermediate block. \\
       $t\in [13.33,16]$  & exec header publishing & The proposer publishes the header \\
        &   &   of the intermediate block with the highest \\
        &   &   number of attestations.\\ \hline
        
    \end{tabular}
    \vspace{11pt}
    \caption{Sequence of Events in a slot pair \cite{2slot}}
    \label{tab:2slot events}
\end{table}

\subsubsection{1-Slot Proposer-Builder Separation} \label{ssec:1_slot_PBS}
As an alternative paradigm to 2-Slot \pbs, \emph{1-Slot Proposer-Builder Separation} was proposed \cite{1slot}. It's removing the dependence on relayers by \say{replacing them with a size $256$ committee of validators.} The block publishing procedure proceeds as follows:
\begin{itemize}
    \item ``A builder splits the payload into $256$ chunks, encrypts each of them with the public key of the respective validator and broadcasts them.''
    \item ``Each committee member validates its chunk. They make a ``pre-attestation'' to the payload's header if valid.''
    \item ``The proposer accepts the payload header with the highest bid and at least 170 signatures.''
    \item ``Upon seeing the proposal, the committee reveals the chunks, and the network reconstructs missing chunks if needed.''
    \item ``Attesters vote on the proposal only if the proposal and the payload are available.''
\end{itemize}


\subsection{Single Unifying Auction for Value Expression (\texttt{SUAVE})}
\label{ssec:SUAVE}
Given the present-day \mev\ situation, the Flashbots team argues that a modification to the transaction life cycle is necessary and should have the following properties:
\begin{itemize}
    \item Users should be able to send transactions to all builders while keeping the transaction details private, and they should receive pre-confirmation privacy. 
    \item These users should receive most of the \mev\ created from their transactions.
    \item The block builders across different chains must integrate openly and permissionless.
\end{itemize}

To achieve the above properties, Flashbots introduced \emph{Single Unifying Auction for Value Expression} (\suave). As mentioned in their article~\cite{suave:futureofmev}, \suave\ is a blockchain with an independent network that can act as a plug-and-play mempool and decentralized block builder for any other blockchain. Parties can collaborate on the expression, execution, and settlement of \emph{preferences} on \suave. Preferences are messages that users sign to indicate desired actions in return for a payment.

\suave\ runs on \emph{\mevm}, a modified version of \evm\ that has precompiled for \mev-related activities, allowing the use of every primitive in the \mev\ supply chain. Using smart contracts to operate on preferences on \mevm\ avoids malicious behaviour (like stealing \mev). Thus, \suave\ decentralizes the current centralized infrastructure of relays, builders, etc.

Flashbots team has published a roadmap for \suave\ and calls all \mev\ stakeholders to experiment with them with test versions of their releases.


\section{Conclusion}
\label{sec:conclusion}

This survey presented the evolution of the \mev\ ecosystem over the last few years. We first briefed about blockchain, consensus mechanisms like PoW and PoS, smart contracts, DeFi, and \mev\ in Section~\ref{sec:introduction} and Section~\ref{sec:preliminaries}. This set up the context to discuss historically how \mev\ extraction happened.

In Section~\ref{sec:mev_landscape_1}, we saw how \mev\ extraction gained attention in Ethereum 1.0 with extractors participating in Priority Gas Auctions. This led to a bloated mempool, increased failed block transactions, higher gas prices, and, thus, a bad user experience. Bundle Auctions introduced by Flashbots took the bidding for \mev\ off-chain while promising \mev\ extractors that either successful bundles get included or none of their bundle transaction get published. A miner could still rob \mev\ by replacing a bidder's bundle with their transactions. This was an even more significant concern in Ethereum 2.0, where the consensus mechanism shifted from PoW to PoS, and the block reward decreased. Block proposers building their own block didn't have much computation to do, unlike in PoW. They could profit by dedicating their computing power to discover and rob \mev\ given that they knew in advance which block they would publish. However, the participants with the best algorithms and computing power and those with the highest funds to stake weren't always the same. This led to Block Auctions, where expert block builders built blocks and auctioned them off to proposers, as discussed in Section~\ref{sec:mev_landscape_2}. Like Bundle Auctions, Block Auctions didn't guarantee builders that the relay or the proposer wouldn't steal their blocks. While a lot of research is being done to develop architectures that provide trustless guarantees to different participants, the majority of the market still engages in block auctions and bundle auctions as of the writing of the survey.

In Section~\ref{sec:private_order_flow}, we discussed the emergence of private order flows, which shielded users from the negative effects of \mev\ extractions, provided faster confirmations, and revert protection. Such exclusive order flow led to centralization, raising the barrier to entry for new searchers and builders. This centralization of builders sabotages the promise of decentralization by blockchains and is a real threat to which the broader blockchain ecosystem is oblivious.

Enshrined \pbs\ and \suave\ are two prominent proposals being researched. \pbs\ modifies the Ethereum protocol to implement \texttt{PBS} in the consensus mechanism. \suave\ introduces a new blockchain with all \mev\ primitives available in its \mevm\ to enable every \mev\ participant to specify their preferences on smart contracts. This research has attracted a lot of researchers from the community, and the latest discussions and proposals can be followed on \href{https://ethresear.ch/}{Ethereum Research's} website. Flashbots publishes their work on their \href{https://writings.flashbots.net/}{Writings website}.

In conclusion, \mev\ extraction, while necessary for the market, threatens to disrupt the ecosystem. More research from different teams is essential to develop trustless solutions in which greedy yet protocol-following agents do not harm the rest of the ecosystem.


\hypersetup{urlcolor=black}
\urlstyle{rm}
\bibliographystyle{ieeetr}
\bibliography{references}
\hypersetup{urlcolor=cyan}
\urlstyle{tt}

\end{document}